\documentclass[]{emulateapj}

\bibpunct[ ]{(}{)}{,}{a}{}{,}

\defcitealias{2002AJ....124.1757S}{SJ02}
\defcitealias{2003Icar..161..174L}{LL03a}
\defcitealias{2003EM&P...92..221L}{LL03b}
\defcitealias{2002NewA....7..359P}{PDR02}
\defcitealias{2000Icar..146..133L}{LRQ}

\newcommand{\myemail}{placerda@strw.leidenuniv.nl}
\newcommand{\tna} {\,\tablenotemark{a}}
\newcommand{\tnb} {\,\tablenotemark{b}}
\newcommand{\tnc} {\,\tablenotemark{c}}
\newcommand{\tnd} {\,\tablenotemark{d}}
\newcommand{\tne} {\,\tablenotemark{e}}
\newcommand{\tnf} {\,\tablenotemark{f}}
\newcommand{\tng} {\,\tablenotemark{g}}
\newcommand{\tnh} {\,\tablenotemark{h}}
\newcommand{\tni} {\,\tablenotemark{i}}
\newcommand{\tnj} {\,\tablenotemark{j}}
\newcommand{\TO} {(19308)\,1996\,\rm{TO}_{66}}
\newcommand{\TOnn} {1996\,\rm{TO}_{66}}
\newcommand{\TS} {1996\,\rm{TS}_{66}}
\newcommand{\SN} {(35671)\,1998\,\rm{SN}_{165}}
\newcommand{\SNnn} {1998\,\rm{SN}_{165}}
\newcommand{\WH} {(19521)\,\rm{Chaos}}

\newcommand{\DF} {(79983)\,1999\,\rm{DF}_{9}}
\newcommand{\DFnn} {1999\,\rm{DF}_{9}}
\newcommand{\RZ} {(66652)\,1999\,\rm{RZ}_{253}}
\newcommand{\RZnn} {1999\,\rm{RZ}_{253}}
\newcommand{\TC} {(47171)\,1999\,\rm{TC}_{36}}
\newcommand{\TCnn} {1999\,\rm{TC}_{36}}
\newcommand{\CM} {(80806)\,2000\,\rm{CM}_{105}}
\newcommand{\CMnn} {2000\,\rm{CM}_{105}}
\newcommand{\EB} {(38628)\,\rm{Huya}}

\newcommand{\CZ} {2001\,\rm{CZ}_{31}}

\shorttitle{Analysis of the Rotational Properties of KBOs}
\shortauthors{Lacerda \& Luu}

\begin{document}


\def\FigA{
  \begin{figure}[]
    \centering
    \includegraphics[width=0.4\textwidth]{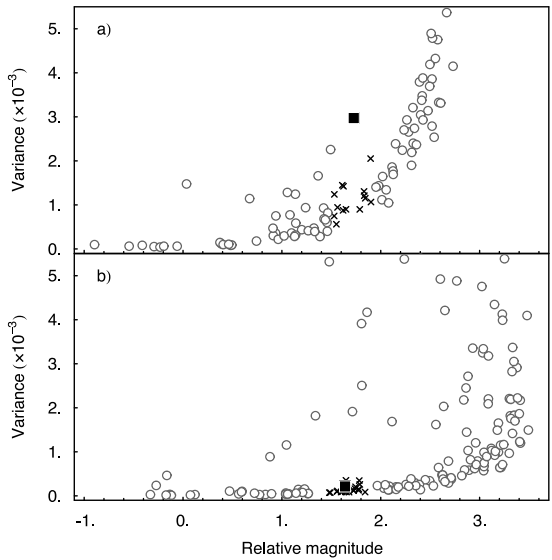}
    \caption[]{Frame-to-frame photometric variances 
      (in magnitudes)
      of all stars (gray circles and black crosses) in the $\SN$ (a)
      and $\EB$ (b) fields, plotted against their relative
      magnitude. The trend of increasing photometric variability with
      increasing magnitude is clear. The intrinsically variable stars
      clearly do not follow this trend, and are located towards the
      upper left region of the plot. The KBOs are shown as black
      squares. $\SN$, in the top panel shows a much larger variability
      than the comparison stars (shown as crosses, see
      Section~\ref{Capitulo4KBOVariability}), while $\EB$ is well
      within the expected variance range, given its magnitude.}
      \label{Fig.VarvsMag}
  \end{figure}
}

\def\FigB{
  \begin{figure*}[]
    \centering
    \includegraphics[width=0.74\textwidth]{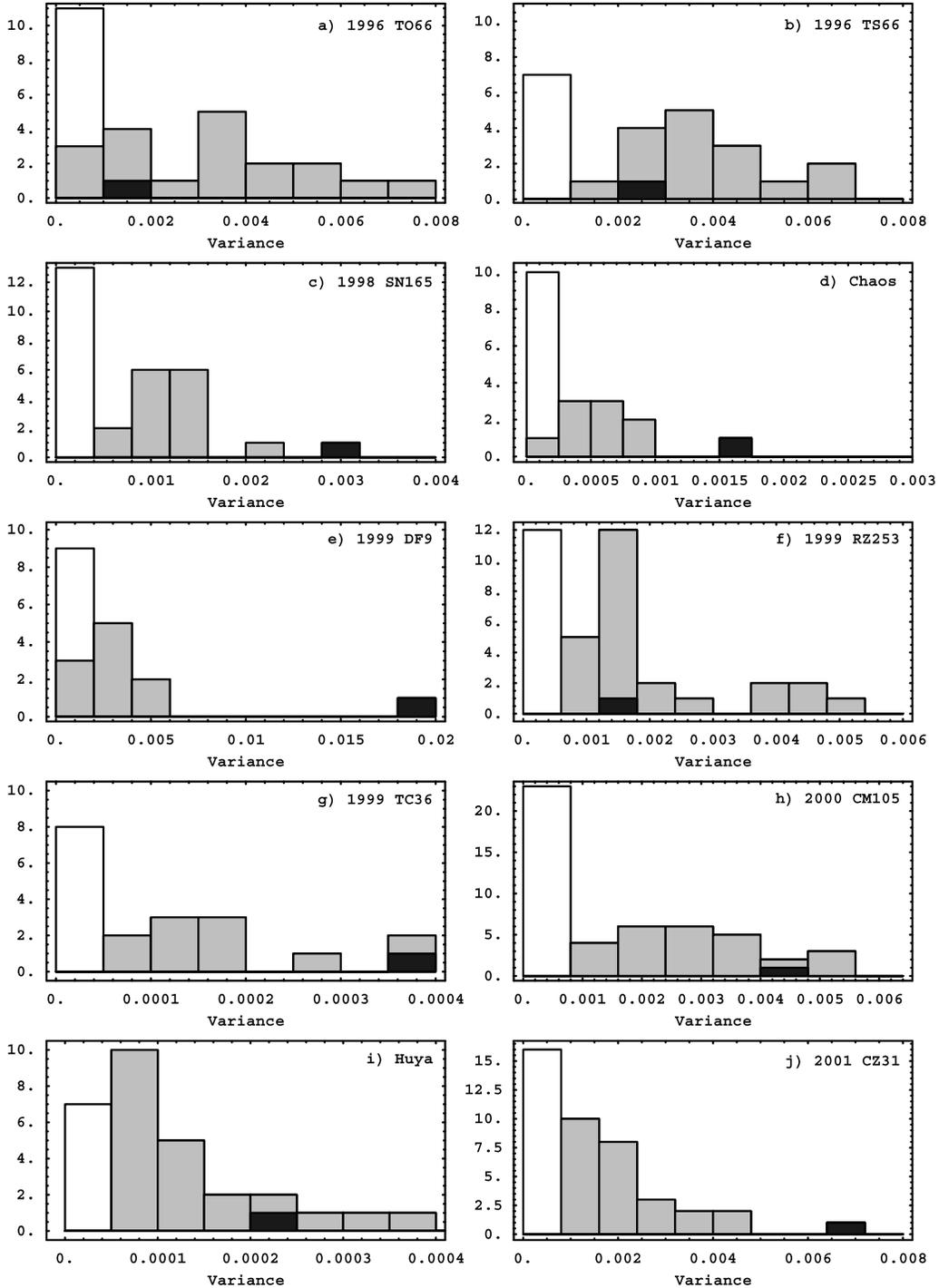}
    \caption[]{Stacked histograms of the
      frame-to-frame variance (in magnitudes) in the optical data 
      on the ``reference''
      stars (in white), ``comparison'' stars (in gray), and the KBO
      (in black). In c), e), and j) the KBO shows significantly more
      variability than the comparison stars, whereas in all other
      cases it falls well within the range of photometric
      uncertainties of the stars of similar brightness.}
      \label{Fig.VarHistograms}
  \end{figure*}
}

\def\FigC{
  \begin{figure}[]
    \centering
    \includegraphics[width=0.4\textwidth]{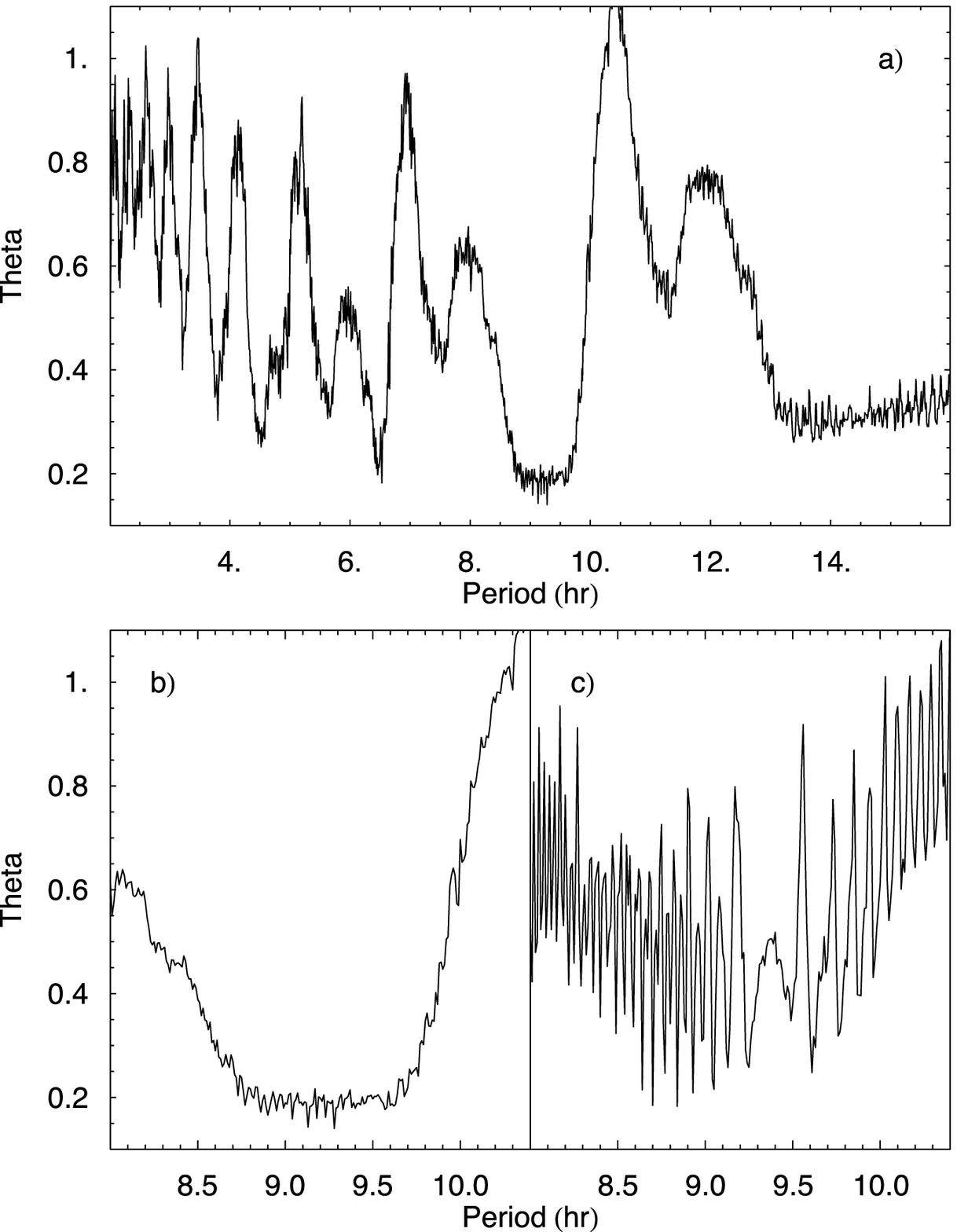}
    \caption[]{Periodogram for the data
      on $\SN$. The lower left panel (b) shows an enlarged section
      near the minimum calculated using only the data published in
      this paper, and the lower right panel (c) shows the same region
      recalculated after adding the PDR02 data.}
  \label{Fig.1998SN165Periodogram}
  \end{figure}
}

\def\FigD{
  \begin{figure}[]
    \centering
    \includegraphics[width=0.4\textwidth]{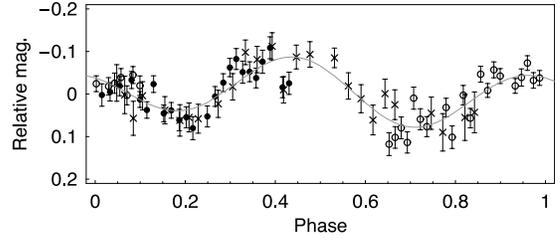}
    \caption[]{Lightcurve of
      $\SN$. The figure represents the data phased with the best fit
      period $P=8.84\,$hr. Different symbols correspond to different
      nights of observation. The gray line is a 2nd order Fourier
      series fit to the data. The PDR02 data are shown as crosses.}
  \label{Fig.1998SN1650884fit}
  \end{figure}
}

\def\FigE{
  \begin{figure}[]
    \centering
    \includegraphics[width=0.4\textwidth]{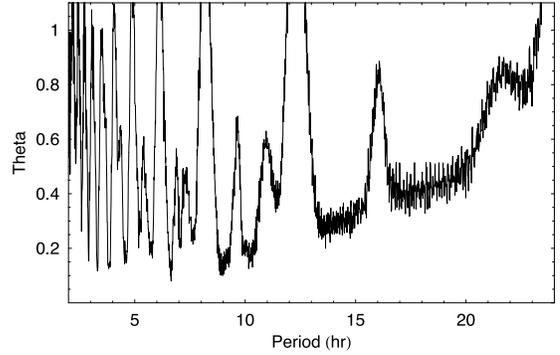}
    \caption[]{Periodogram for the $\DF$
      data. The minimum corresponds to a lightcurve period
      $P=6.65\,$hr.}
  \label{Fig.1999DF9.periodogram}
  \end{figure}
}

\def\FigF{
  \begin{figure}[]
    \centering
    \includegraphics[width=0.4\textwidth]{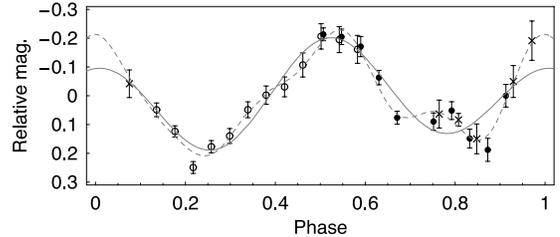}
    \caption[]{Same as
      Fig.~\ref{Fig.1998SN1650884fit} for KBO $\DF$. The best fit
      period is $P=6.65\,$hr. The lines represent a 2nd order (solid
      line) and 5th (dashed line) order Fourier series fits to the
      data. Normalized $\chi^2$ values of the fits are 2.8 and 1.3
      respectively.}
  \label{Fig.1999DF90665fit}
  \end{figure}
}

\def\FigG{
  \begin{figure}[]
    \centering
    \includegraphics[width=0.4\textwidth]{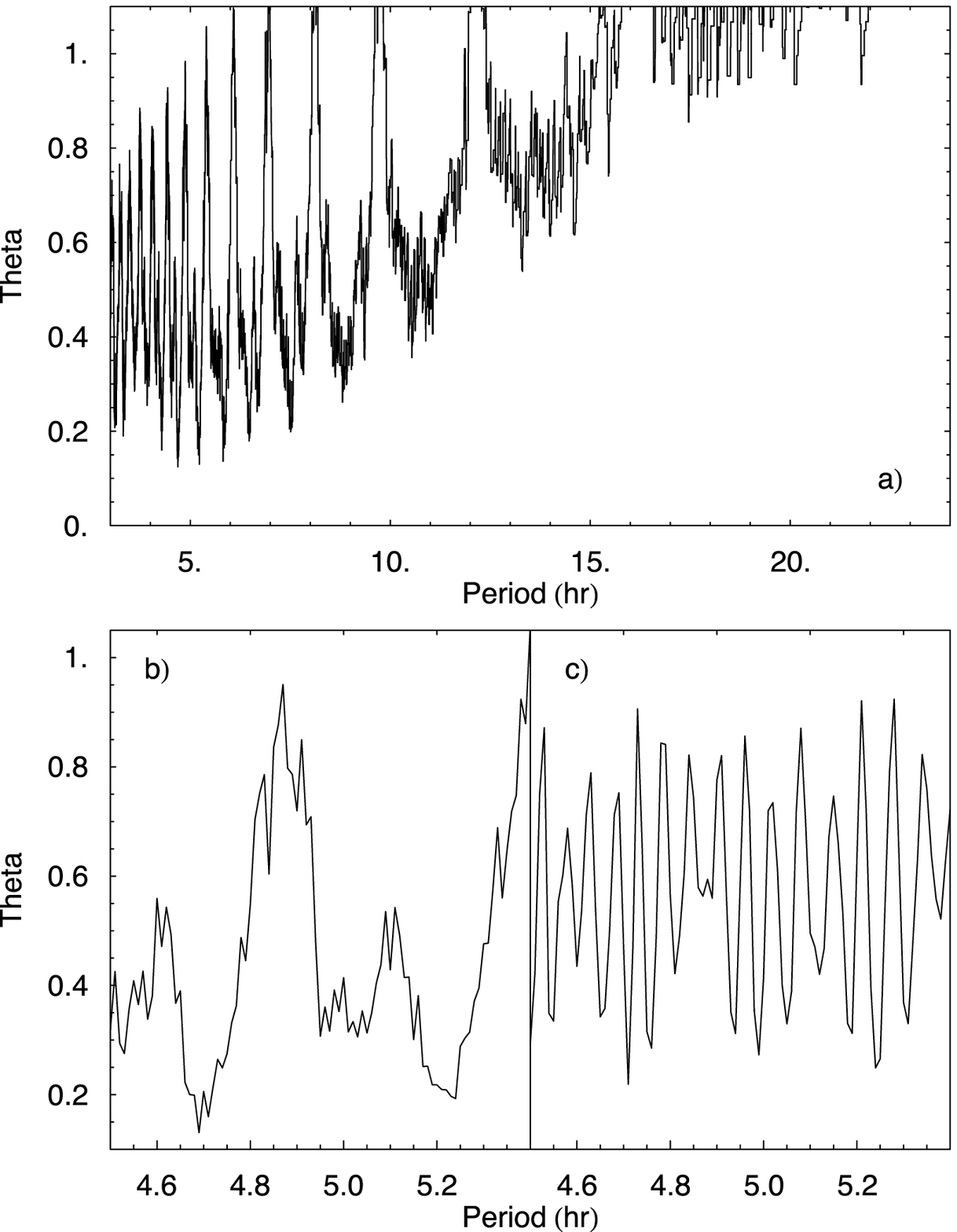}
    \caption[]{Periodogram for the $\CZ$
      data. The lower left panel (b) shows an enlarged section near
      the minimum calculated using only the data published in this
      paper, and the lower right panel (c) shows the same region
      recalculated after adding the SJ02 data.}
  \label{Fig.2001CZ31.periodogram}
  \end{figure}
}

\def\FigH{
  \begin{figure}[]
    \centering
    \includegraphics[width=0.4\textwidth]{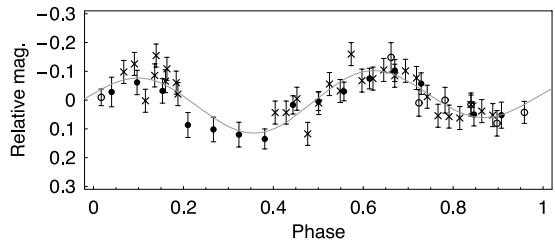}
    \caption[]{Same as
      Fig.~\ref{Fig.1998SN1650884fit} for KBO $\CZ$. The data are
      phased with period $P=4.71\,$hr. The points represented by
      crosses are taken from SJ02.}
  \label{Fig.2001CZ31.0471fit}
  \end{figure}
}

\def\FigI{
  \begin{figure*}[]
    \centering
    \includegraphics[width=0.7\textwidth]{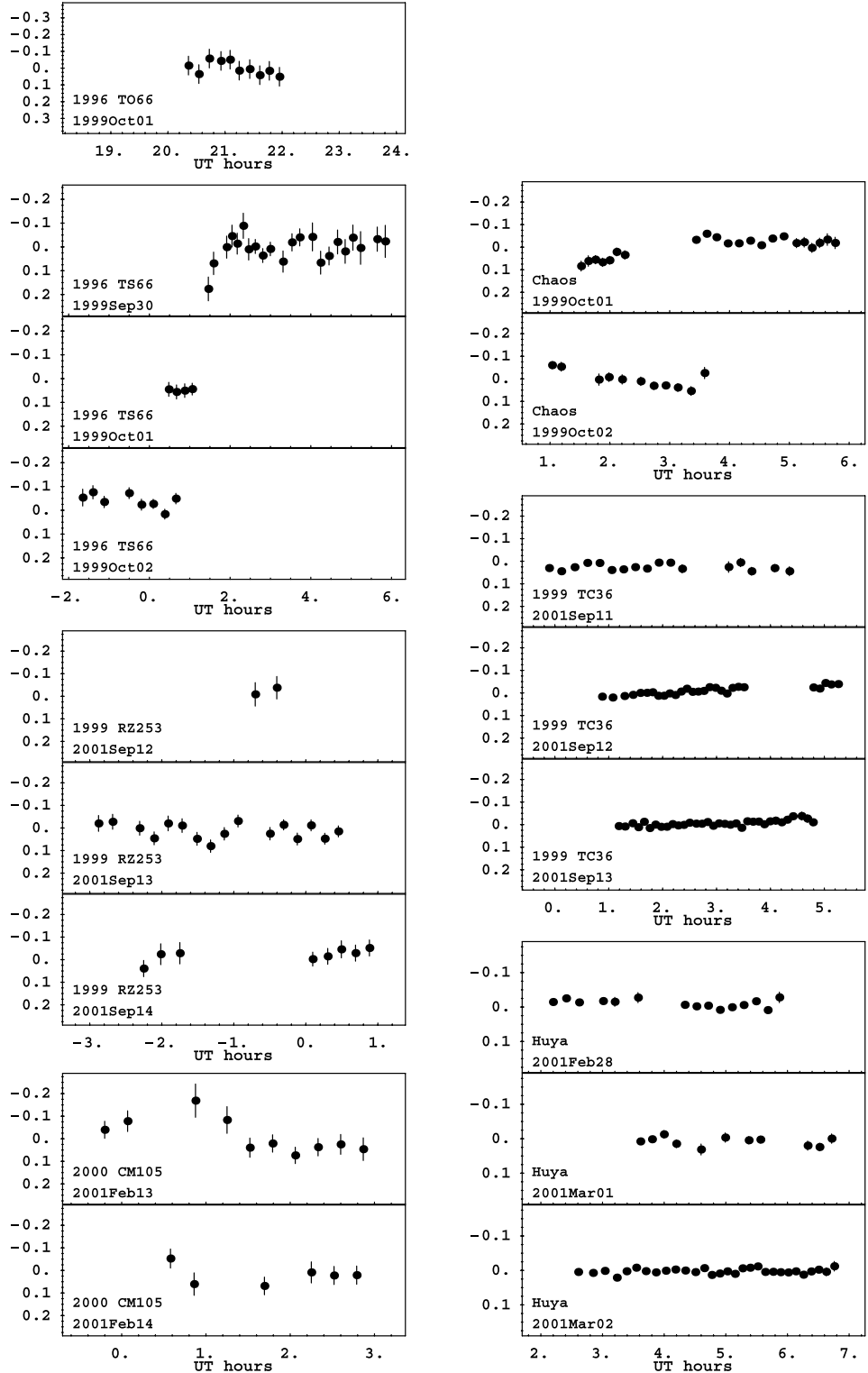}
    \caption[]{The ``flat'' lightcurves are
      shown. The respective amplitudes are within the photometric
      uncertainties.}
  \label{Fig.FlatLCurves}
  \end{figure*}
}

\def\FigJ{
  \begin{figure}[]
    \centering
    \includegraphics[width=0.4\textwidth]{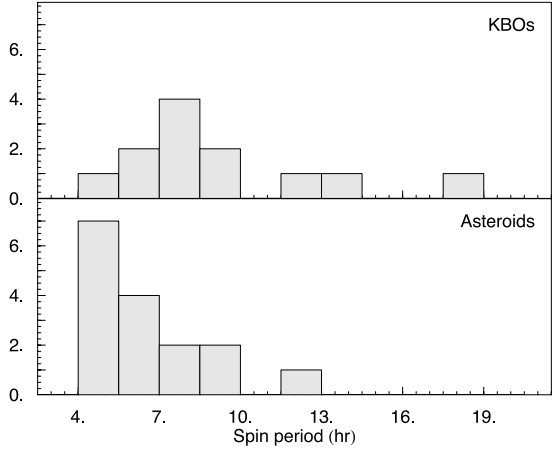}
    \caption[]{Histograms of the spin
      periods of KBOs (upper panel) and main belt asteroids (lower
      panel) satisfying $D>200\,$km, $\Delta m \geq 0.15\,$mag,
      $P<20\,$hr. The mean rotational periods of KBOs and MBAs are
      9.23$\,$hr and 6.48$\,$hr, respectively. The $y$-axis in both
      cases indicates the number of objects in each range of spin
      periods.}
  \label{Fig.PeriodDistrib}
  \end{figure}
}

\def\FigK{
  \begin{figure}[]
    \centering
    \includegraphics[width=0.4\textwidth]{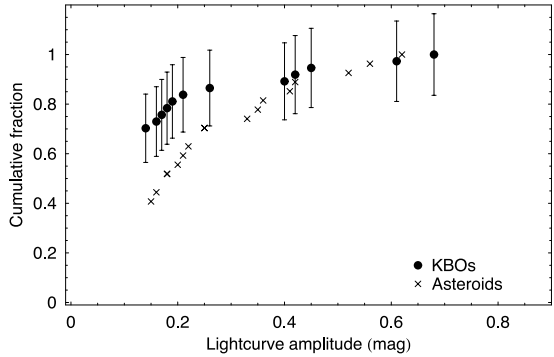}
    \caption[]{Cumulative
      distribution of lightcurve amplitude for KBOs (circles) and
      asteroids (crosses) larger than $200\,$km. We plot only KBOs for
      which a period has been determined. KBO 2001$\,QG_{298}$,
      thought to be a contact binary \citep{2004AJ....127.3023S}, is
      not plotted although it may be considered an extreme case of
      elongation.}
  \label{Fig.CumulAmpl}
  \end{figure}
}

\def\FigL{
  \begin{figure}[]
    \centering
    \includegraphics[width=0.4\textwidth]{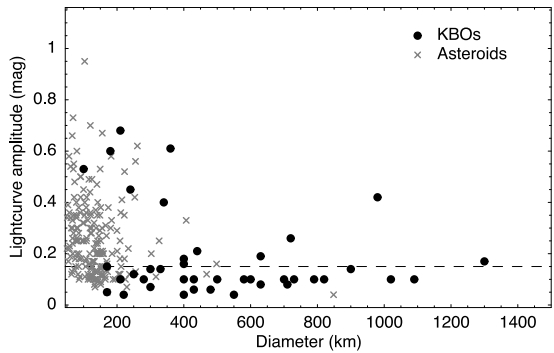}
    \caption[]{Lightcurve amplitudes of KBOs
      (black circles) and main belt asteroids (gray crosses) plotted
      against object size.}
  \label{Fig.AmplvsSize}
  \end{figure}
}

\def\FigM{
  \begin{figure}[]
    \centering
    \includegraphics[width=0.4\textwidth]{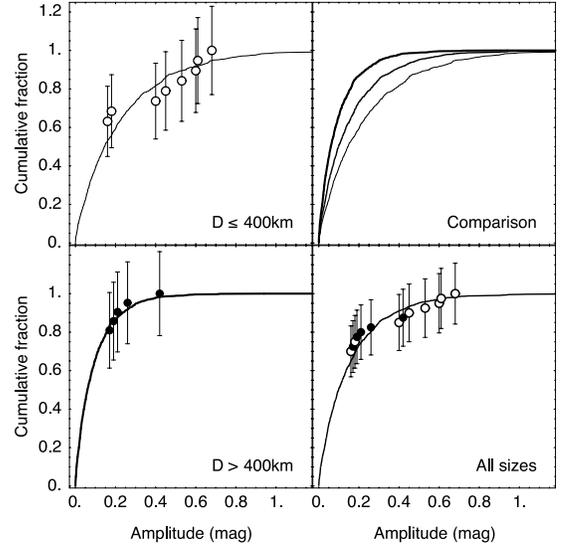}
    \caption[]{Observed
      cumulative lightcurve amplitude distributions of KBOs (black
      circles) with diameter smaller than $400\,$km (upper left
      panel), larger than $400\,$km (lower left panel), and of all the
      sample (lower right panel) are shown as black circles. Error
      bars are Poissonian. The best fit power-law shape distributions
      of the form $f(\tilde{a})\,{\rm d}\tilde{a}=\tilde{a}^{-q}\,{\rm
      d}\tilde{a}$ were converted to amplitude distributions using a
      Monte Carlo technique (see text for details), and are shown as
      solid lines. The best fit $q$'s are listed in
      Table~\ref{Table.BestFitShapeDist}.}
  \label{Fig.CumulDistrFit}
  \end{figure}
}

\def\FigN{
  \begin{figure}[]
    \centering
    \includegraphics[width=0.4\textwidth]{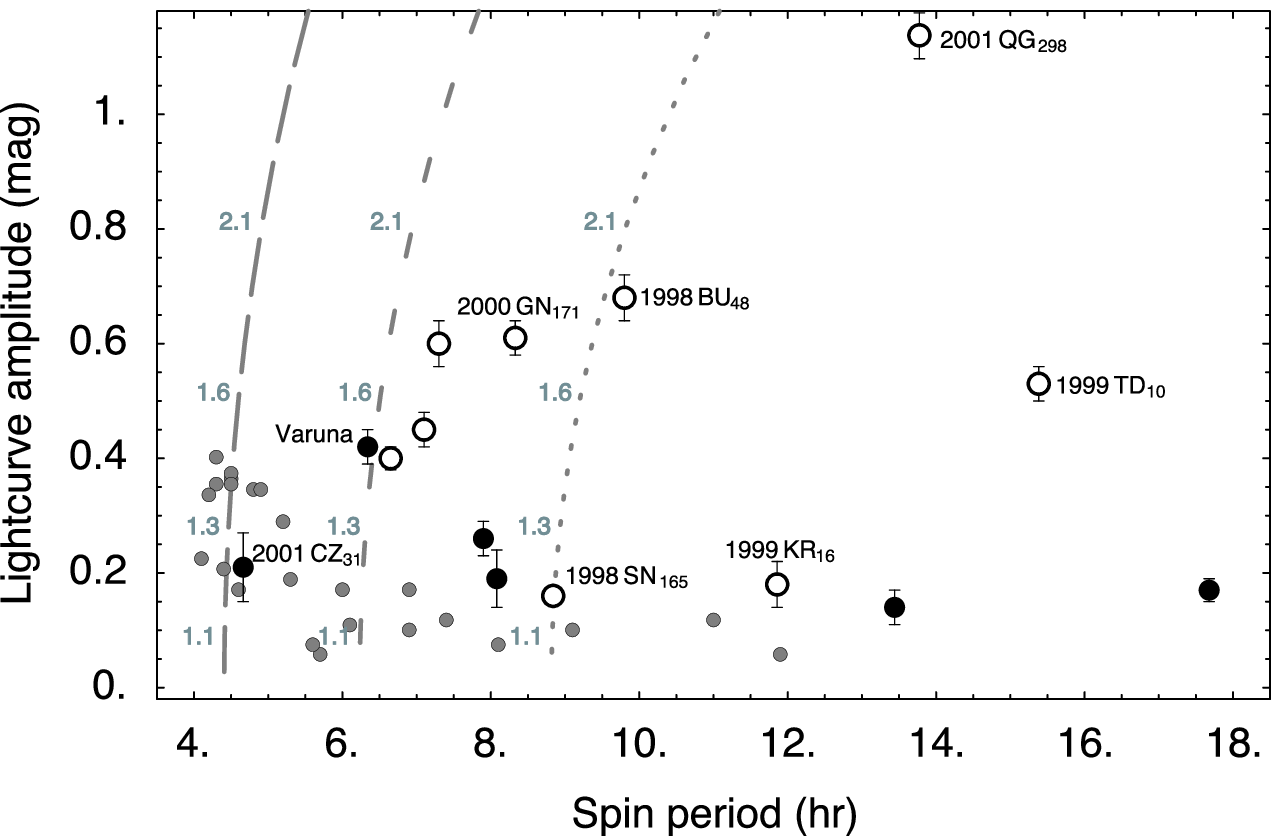}
    \caption[]{Lightcurve amplitudes versus
      spin periods of KBOs. The black filled and open circles
      represent objects larger and smaller than 400$\,$km,
      respectively. The smaller gray circles show the results of
      numerical simulations of ``rubble-pile'' collisions
      \citep{2000Icar..146..133L}. The lines represent the locus of
      rotating ellipsoidal figures of hydrostatic equilibrium with
      densities $\rho=500\,{\rm kg}\,{\rm m}^{-3}$ (dotted line),
      $\rho=1000\,{\rm kg}\,{\rm m}^{-3}$ (shorter dashes) and
      $\rho=2000\,{\rm kg}\,{\rm m}^{-3}$ (longer dashes). Shown in grey
      next to the lines are the axis ratios, $a/b$, of the ellipsoidal 
      solutions. Both the \citet{2000Icar..146..133L} results and the 
      figures of equilibrium assume equator-on observing geometry and 
      therefore represent upper limits.}
  \label{Fig.PvsAmpl}
  \end{figure}
}

\title{Analysis of the Rotational Properties of Kuiper Belt Objects}

\author{Pedro Lacerda\altaffilmark{1,2}}
\affil{GAUC, Departamento de Matem\'atica, Lg. D. Dinis, 3000 Coimbra,
Portugal}
\email{\myemail}

\author{Jane Luu}
\affil{MIT Lincoln Laboratory, 244 Wood Street, Lexington, MA 02420, USA}

\altaffiltext{1}{Institute for Astronomy, University of Hawaii, 2680 Woodlawn
Drive, Honolulu, HI 96822}

\altaffiltext{2}{Leiden Observatory, Postbus 9513, NL-2300 RA Leiden,
Netherlands}

\begin{abstract}
  We use optical data on 10 Kuiper Belt objects (KBOs) to investigate
  their rotational properties. Of the 10, three (30\%) exhibit light
  variations with amplitude $\Delta m \ge 0.15\,$mag, and 1 out of 10
  (10\%) has $\Delta m \ge 0.40\,$mag, which is in good agreement with
  previous surveys.  These data, in combination with the existing
  database, are used to discuss the rotational periods, shapes, and
  densities of Kuiper Belt objects. We find that, in the sampled size
  range, Kuiper Belt objects have a higher fraction of low amplitude
  lightcurves and rotate slower than main belt asteroids. The data
  also show that the rotational properties and the shapes of KBOs
  depend on size. If we split the database of KBO rotational
  properties into two size ranges with diameter {\em larger} and {\em
    smaller} than $400\,$km, we find that: (1) the mean lightcurve
  amplitudes of the two groups are different with 98.5\% confidence,
  (2) the corresponding power-law shape distributions seem to be 
  different, although the existing data are too sparse to render this
  difference significant, and (3) the two groups occupy different 
  regions on a {\em spin period} vs. {\em lightcurve amplitude} diagram.  
  These differences are interpreted in the context of KBO collisional 
  evolution.
\end{abstract}

\keywords{Kuiper Belt objects --- minor planets, asteroids --- solar
  system: general}

\section{Introduction}

The Kuiper Belt (KB) is an assembly of mostly small icy objects, orbiting the
Sun beyond Neptune. Kuiper Belt objects (KBOs) are likely to be
remnants of outer solar system planetesimals
\citep{1993Natur.362..730J}.  Their physical, chemical, and dynamical
properties should therefore provide valuable information regarding
both the environment and the physical processes responsible for planet
formation.

At the time of writing, roughly 1000 KBOs are known, half of which
have been followed for more than one opposition. A total of $\approx
10^{5}$ objects larger than 50~km are thought to orbit the Sun beyond
Neptune \citep{2000prpl.conf.1201J}.  Studies of KB orbits have
revealed an intricate dynamical structure, with signatures of
interactions with Neptune (Malhotra 1995). The size distribution
follows a differential power-law of index $q=4$ for bodies $\gtrsim
50\,$km \citep{2001AJ....122..457T}, becoming slightly shallower at
smaller sizes \citep{2004AJ....128.1364B}.

KBO colours show a large diversity, from slightly blue to very red
\citep{1996AJ....112.2310L,2000Natur.407..979T,2001AJ....122.2099J}, and seem
to correlate with inclination and/or perihelion distance
\citep[e.g.,][]{2001AJ....122.2099J,2002AJ....124.2279D, 2002ApJ...566L.125T}.
The few low-resolution optical and near-IR KBO spectra are mostly featureless,
with the exception of a weak 2$\,\mu$m water ice absorption line present in
some of them \citep{1999ApJ...519L.101B,2001AJ....122.2099J}, and strong 
methane absorption on 2003$\,$UB$_{313}$ \citep{2003UB313}.

About 4\% of known KBOs are binaries with separations larger than
0\farcs15 \citep{2002AJ....124.3424N}. All the observed binaries have
primary-to-secondary mass ratios $\approx$ 1. Two binary creation
models have been proposed. \citet{2002Icar..160..212W} favours the
idea that binaries form in three-body encounters. This model requires
a 100 times denser Kuiper Belt at the epoch of binary formation, and
predicts a higher abundance of large separation binaries. An
alternative scenario \citep{2002Natur.420..643G}, in which the energy
needed to bind the orbits of two approaching bodies is drawn from the
surrounding swarm of smaller objects, also requires a much higher
density of KBOs than the present, but it predicts a larger fraction of
close binaries. Recently, \citet{2004AJ....127.3023S} have shown
evidence that $2001\,\rm{QG}_{298}$ could be a close or contact binary
KBO, and estimated the fraction of similar objects in the Belt to be
$\sim10\%$--$20\%$.

Other physical properties of KBOs, such as their shapes, densities,
and albedos, are still poorly constrained.  This is mainly because
KBOs are extremely faint, with mean apparent red magnitude $m_R\sim$23
\citep{2001AJ....122.2740T}.

The study of KBO rotational properties through time-series broadband
optical photometry has proved to be the most successful technique to
date to investigate some of these physical properties.  Light
variations of KBOs are believed to be caused mainly by their
aspherical shape: as KBOs rotate in space, their projected
cross-sections change, resulting in periodic brightness variations.

One of the best examples to date of a KBO lightcurve -- and what can
be learned from it -- is that of (20000)$\,$Varuna
\citep{2002AJ....123.2110J}. The authors explain the lightcurve of
(20000)$\,$Varuna as a consequence of its elongated shape (axes ratio,
$a/b\sim 1.5$). They further argue that the object is centripetally
deformed by rotation because of its low density, ``rubble pile''
structure. The term ``rubble pile'' is generally used to refer to
gravitationally bound aggregates of smaller fragments. The existence
of rubble piles is thought to be due to continuing mutual collisions
throughout the age of the solar system, which gradually fracture the
interiors of objects. Rotating rubble piles can adjust their shapes to
balance centripetal acceleration and self-gravity. The resulting
equilibrium shapes have been studied in the extreme case of fluid
bodies, and depend on the body's density and spin rate
\citep{1969efe..book.....C}.
  
\citeauthor{2003Icar..161..174L} (\citeyear{2003Icar..161..174L},
hereafter \citetalias{2003Icar..161..174L}) showed that under
reasonable assumptions the fraction of KBOs with detectable
lightcurves can be used to constrain the shape distribution of these
objects. A follow-up (\citealt{2003EM&P...92..221L}, hereafter
\citetalias{2003EM&P...92..221L}) on this work, using a database of
lightcurve properties of 33 KBOs
\citep[]{2002AJ....124.1757S,2003EM&P...92..207S}, shows that although
most Kuiper Belt objects ($\sim85\%$) have shapes that are close to
spherical ($a/b\leq1.5$) there is a significant fraction ($\sim12\%$)
with highly aspherical shapes ($a/b\geq1.7$).

In this paper we use optical data on 10 KBOs to investigate the
amplitudes and periods of their lightcurves. These data are used in
combination with the existing database to investigate the
distributions of KBO spin periods and shapes. We discuss their
implications for the inner structure and collisional evolution of
objects in the Kuiper Belt.

\section{Observations and Photometry}
\label{Capitulo4Sec.Observations}

\FigA

\FigB

  \begin{deluxetable*}{r@{ }lcccccccccc}
    \tabletypesize{\scriptsize}
    \tablecaption{Observing Conditions and Geometry \label{Table.ObsCond}}
    \tablewidth{0pt}
    \tablehead{
      \multicolumn{2}{c}{Object Designation} & \colhead{ObsDate\tna} & 
      \colhead{Tel.\tnb} & \colhead{Seeing\tnc} & \colhead{MvtRt\tnd} & 
      \colhead{ITime\tne} & \colhead{RA\tnf} & \colhead{dec\tng} & 
      \colhead{$R$\tnh} & \colhead{$\Delta$\tni} & \colhead{$\alpha$\tnj} \\
      \multicolumn{4}{c}{ } & \colhead{[\arcsec]} & \colhead{[\arcsec/hr]} & 
      \colhead{[sec]} & \colhead{[hhmmss]} & 
      \colhead{[\arcdeg \arcmin \arcsec]} & \colhead{[AU]} & \colhead{[AU]} & 
      \colhead{[deg]}
      }
      \startdata
(19308) &$\TOnn$& 99-Oct-01 & WHT & 1.8 & 2.89 & 500         & 23 59 46 &   +03 36 42 & 45.950 & 44.958 & 0.1594\\
        & $\TS$ & 99-Sep-30 & WHT & 1.3 & 2.62 & 400,600     & 02 26 06 &   +21 41 03 & 38.778 & 37.957 & 0.8619\\
        & $\TS$ & 99-Oct-01 & WHT & 1.1 & 2.67 & 600         & 02 26 02 &   +21 40 50 & 38.778 & 37.948 & 0.8436\\
        & $\TS$ & 99-Oct-02 & WHT & 1.5 & 2.70 & 600,900     & 02 25 58 &   +21 40 35 & 38.778 & 37.939 & 0.8225\\
(35671) &$\SNnn$& 99-Sep-29 & WHT & 1.5 & 3.24 & 360,400     & 23 32 46 & $-$01 18 15 & 38.202 & 37.226 & 0.3341\\
(35671) &$\SNnn$& 99-Sep-30 & WHT & 1.4 & 3.22 & 360         & 23 32 41 & $-$01 18 47 & 38.202 & 37.230 & 0.3594\\
(19521) & Chaos & 99-Oct-01 & WHT & 1.0 & 1.75 & 360,400,600 & 03 44 37 &   +21 30 58 & 42.399 & 41.766 & 1.0616\\
(19521) & Chaos & 99-Oct-02 & WHT & 1.5 & 1.79 & 400,600     & 03 44 34 &   +21 30 54 & 42.399 & 41.755 & 1.0484\\
(79983) &$\DFnn$& 01-Feb-13 & WHT & 1.7 & 3.19 & 900         & 10 27 04 &   +09 45 16 & 39.782 & 38.818 & 0.3124\\
(79983) &$\DFnn$& 01-Feb-14 & WHT & 1.6 & 3.21 & 900         & 10 26 50 &   +09 46 25 & 39.783 & 38.808 & 0.2436\\
(79983) &$\DFnn$& 01-Feb-15 & WHT & 1.4 & 3.22 & 900         & 10 26 46 &   +09 46 50 & 39.783 & 38.806 & 0.2183\\
(80806) &$\CMnn$& 01-Feb-11 & WHT & 1.5 & 3.14 & 600,900     & 09 18 48 &   +19 41 59 & 41.753 & 40.774 & 0.1687\\
(80806) &$\CMnn$& 01-Feb-13 & WHT & 1.4 & 3.12 & 900         & 09 18 39 &   +19 42 40 & 41.753 & 40.778 & 0.2084\\
(80806) &$\CMnn$& 01-Feb-14 & WHT & 1.5 & 3.11 & 900         & 09 18 34 &   +19 43 02 & 41.753 & 40.781 & 0.2303\\
(66652) &$\RZnn$& 01-Sep-11 & INT & 1.9 & 2.86 & 600         & 22 02 57 & $-$12 31 06 & 40.963 & 40.021 & 0.4959\\
(66652) &$\RZnn$& 01-Sep-12 & INT & 1.4 & 2.84 & 600         & 22 02 53 & $-$12 31 26 & 40.963 & 40.026 & 0.5156\\
(66652) &$\RZnn$& 01-Sep-13 & INT & 1.8 & 2.82 & 600         & 22 02 49 & $-$12 31 49 & 40.963 & 40.033 & 0.5381\\
(47171) &$\TCnn$& 01-Sep-11 & INT & 1.9 & 3.85 & 600         & 00 16 49 & $-$07 34 59 & 31.416 & 30.440 & 0.4605\\
(47171) &$\TCnn$& 01-Sep-12 & INT & 1.4 & 3.86 & 900         & 00 16 45 & $-$07 35 33 & 31.416 & 30.437 & 0.4359\\
(47171) &$\TCnn$& 01-Sep-13 & INT & 1.8 & 3.88 & 900         & 00 16 39 & $-$07 36 13 & 31.416 & 30.434 & 0.4122\\
(38628) & Huya  & 01-Feb-28 & INT & 1.5 & 2.91 & 600         & 13 31 13 & $-$00 39 04 & 29.769 & 29.021 & 1.2725\\
(38628) & Huya  & 01-Mar-01 & INT & 1.8 & 2.97 & 360         & 13 31 09 & $-$00 38 23 & 29.768 & 29.009 & 1.2479\\
(38628) & Huya  & 01-Mar-03 & INT & 1.5 & 3.08 & 360         & 13 31 01 & $-$00 36 59 & 29.767 & 28.987 & 1.1976\\
        & $\CZ$ & 01-Mar-01 & INT & 1.3 & 2.72 & 600,900     & 09 00 03 &   +16 29 23 & 41.394 & 40.522 & 0.6525\\
        & $\CZ$ & 01-Mar-03 & INT & 1.4 & 2.65 & 600,900     & 08 59 54 &   +16 30 04 & 41.394 & 40.539 & 0.6954\\
      \enddata
      \tablenotetext{a}{UT date of observation}
      \tablenotetext{b}{Telescope used for observations}
      \tablenotetext{c}{Average seeing of the data [\arcsec]}
      \tablenotetext{d}{Average rate of motion of KBO [\arcsec/hr]}
      \tablenotetext{e}{Integration times used}
      \tablenotetext{f}{Right ascension}
      \tablenotetext{g}{Declination}
      \tablenotetext{h}{KBO--Sun distance}
      \tablenotetext{i}{KBO--Earth distance}
      \tablenotetext{j}{Phase angle (Sun--Object--Earth angle) of observation}
  \end{deluxetable*}

  \begin{deluxetable}{r@{ }lccrcc}
    \tabletypesize{\scriptsize}
    \tablecaption{Properties of Observed KBOs \label{ParsObsKBO}}
    \tablewidth{0pt}
    \tablehead{
      \multicolumn{2}{c}{Object Designation} & \colhead{Class\tna} & 
      \colhead{$H$\tnb} & \colhead{$i$\tnc} & \colhead{$e$\tnd} & 
      \colhead{$a$\tne} \\ \multicolumn{3}{c}{ } & \colhead{[mag]} & 
      \colhead{[deg]} & \colhead{ } & \colhead{[AU]}
    }
    \startdata
(19308) &$\TOnn$& C     & 4.5 &27.50 & 0.12 & 43.20 \\
        & $\TS$ & C     & 6.4 & 7.30 & 0.13 & 44.00 \\
(35671) &$\SNnn$& C\tnf & 5.8 & 4.60 & 0.05 & 37.80 \\
(19521) & Chaos & C     & 4.9 &12.00 & 0.11 & 45.90 \\
(79983) &$\DFnn$& C     & 6.1 & 9.80 & 0.15 & 46.80 \\
(80806) &$\CMnn$& C     & 6.2 & 3.80 & 0.07 & 42.50 \\
(66652) &$\RZnn$& C     & 5.9 & 0.60 & 0.09 & 43.60 \\
(47171) &$\TCnn$& Pb    & 4.9 & 8.40 & 0.22 & 39.30 \\
(38628) & Huya  & P     & 4.7 &15.50 & 0.28 & 39.50 \\
        & $\CZ$ & C     & 5.4 &10.20 & 0.12 & 45.60 \\
    \enddata
    \tablenotetext{a}{Dynamical class (C = Classical KBO, P = Plutino, b = binary KBO)}
    \tablenotetext{b}{Absolute magnitude}
    \tablenotetext{c}{Orbital inclination}
    \tablenotetext{d}{Orbital eccentricity}
    \tablenotetext{e}{Semi-major axis}
    \tablenotetext{f}{This object as a classical-type inclination and
      eccentricity but its semi-major axis is much smaller than
      for other classical KBOs}
  \end{deluxetable}

We collected time-series optical data on 10 KBOs at the Isaac Newton
2.5m (INT) and William Herschel 4m (WHT) telescopes. The INT Wide
Field Camera (WFC) is a mosaic of 4 EEV 2048$\times$4096 CCDs, each
with a pixel scale of 0\farcs33/pixel and spanning approximately
11\farcm3$\times$22\farcm5 in the plane of the sky. The targets are
imaged through a Johnson R filter. The WHT prime focus camera consists
of 2 EEV 2048$\times$4096 CCDs with a pixel scale of 0\farcs24/pixel,
and covers a sky-projected area of
2$\times$8\farcm2$\times$16\farcm4. With this camera we used a Harris
R filter. The seeing for the whole set of observations ranged from 1.0
to 1.9\arcsec FWHM. We tracked both telescopes at sidereal rate and
kept integration times for each object sufficiently short to avoid
errors in the photometry due to trailing effects (see
Table~\ref{Table.ObsCond}). No light travel time corrections have been
made.

We reduced the data using standard techniques. The sky background in
the flat-fielded images shows variations of less than 1\% across the
chip. Background variations between consecutive nights were less than
5\% for most of the data. Cosmic rays were removed with the package
LA-Cosmic \citep{2001PASP..113.1420V}.

We performed aperture photometry on all objects in the field using the
SExtractor software package \citep{1996A&AS..117..393B}. This software
performs circular aperture measurements on each object in a frame, and
puts out a catalog of both the magnitudes and the associated
errors. Below we describe how we obtained a better estimate of the
errors. We used apertures ranging from 1.5 to 2.0 times the FWHM for
each frame and selected the aperture that maximized
signal-to-noise. An extra aperture of 5 FWHMs was used to look for
possible seeing dependent trends in our photometry. The catalogs were
matched by selecting only the sources that are present in all
frames. The slow movement of KBOs from night to night allows us to
successfully match a large number of sources in consecutive nights.
We discarded all saturated sources as well as those identified to be
galaxies.

The KBO lightcurves were obtained from differential photometry with
respect to the brightest non-variable field stars. An average of the
magnitudes of the brightest stars (the "reference" stars) provides a
reference for differential photometry in each frame. This method
allows for small amplitude brightness variations to be detected even
under non-photometric conditions.
  
The uncertainty in the relative photometry was calculated from the
scatter in the photometry of field stars that are similar to
the KBOs in brightness (the "comparison" stars, see
Fig.\ref{Fig.VarvsMag}). This error estimate is more robust than the
errors provided by SExtractor (see below), and was used to verify the
accuracy of the latter.  This procedure resulted in consistent time
series brightness data for $\sim100$ objects (KBO + field stars) in a
time span of 2--3 consecutive nights.

We observed Landolt standard stars whenever conditions were
photometric, and used them to calibrate the zero point of the
magnitude scale. The extinction coefficient was obtained from the
reference stars.

Since not all nights were photometric the lightcurves are presented as
variations with respect to the mean brightness. These yield the
correct amplitudes and periods of the lightcurves but do not provide
their absolute magnitudes.
  
The orbital parameters and other properties of the observed KBOs are
given in Table~\ref{ParsObsKBO}.  Tables~\ref{Table.1996TO66AbsPhot},
\ref{Table.1996TS66AbsPhot}, \ref{Table.1998SN165AbsPhot}, and
\ref{Table.1998WH24AbsPhot} list the absolute $R$-magnitude
photometric measurements obtained for $\TO$, $\TS$, $\SN$, and $\WH$,
respectively. Tables~\ref{Table.1999DF9RelPhot} and 
\ref{Table.2001CZ31RelPhot} list the mean-subtracted $R$-band data for 
$\DF$ and $\CZ$.

  \begin{deluxetable}{ccc}
    \tabletypesize{\scriptsize}
    \tablecaption{Photometric measurements of $\TO$
    \label{Table.1996TO66AbsPhot}}
    \tablewidth{0pt}
    \tablehead{
	\colhead{ } & \colhead{ } & \colhead{$m_R$\tnc} \\
	\colhead{UT Date\tna} & \colhead{Julian Date\tnb} & \colhead{[mag]}
    }
    \startdata
    1999 Oct 1.84831 &  2451453.34831  & 21.24 $\pm$ 0.07 \\
    1999 Oct 1.85590 &  2451453.35590  & 21.30 $\pm$ 0.07 \\
    1999 Oct 1.86352 &  2451453.36352  & 21.20 $\pm$ 0.07 \\
    1999 Oct 1.87201 &  2451453.37201  & 21.22 $\pm$ 0.07 \\
    1999 Oct 1.87867 &  2451453.37867  & 21.21 $\pm$ 0.07 \\
    1999 Oct 1.88532 &  2451453.38532  & 21.28 $\pm$ 0.07 \\
    1999 Oct 1.89302 &  2451453.39302  & 21.27 $\pm$ 0.06 \\
    1999 Oct 1.90034 &  2451453.40034  & 21.30 $\pm$ 0.06 \\
    1999 Oct 1.90730 &  2451453.40730  & 21.28 $\pm$ 0.06 \\
    1999 Oct 1.91470 &  2451453.41470  & 21.31 $\pm$ 0.06 \\    
    \enddata
    \tablenotetext{a}{UT date at the start of the exposure}
    \tablenotetext{b}{Julian date at the start of the exposure}
    \tablenotetext{c}{Apparent red magnitude; errors include
      uncertainties in relative and absolute photometry added
      quadratically}
  \end{deluxetable}

  \begin{deluxetable}{ccc}
    \tabletypesize{\scriptsize}
    \tablecaption{Photometric measurements of $\TS$
    \label{Table.1996TS66AbsPhot}}
    \tablewidth{0pt}
    \tablehead{
	\colhead{ } & \colhead{ } & \colhead{$m_R$\tnc} \\
	\colhead{UT Date\tna} & \colhead{Julian Date\tnb} & \colhead{[mag]}
    }
    \startdata
    1999 Sep 30.06087 & 2451451.56087 & 21.94 $\pm$ 0.07 \\
    1999 Sep 30.06628 & 2451451.56628 & 21.83 $\pm$ 0.07 \\
    1999 Sep 30.07979 & 2451451.57979 & 21.76 $\pm$ 0.07 \\
    1999 Sep 30.08529 & 2451451.58529 & 21.71 $\pm$ 0.07 \\
    1999 Sep 30.09068 & 2451451.59068 & 21.75 $\pm$ 0.07 \\
    1999 Sep 30.09695 & 2451451.59695 & 21.67 $\pm$ 0.07 \\
    1999 Sep 30.01250 & 2451451.60250 & 21.77 $\pm$ 0.07 \\
    1999 Sep 30.10936 & 2451451.60936 & 21.76 $\pm$ 0.06 \\
    1999 Sep 30.11705 & 2451451.61705 & 21.80 $\pm$ 0.06 \\
    1999 Sep 30.12486 & 2451451.62486 & 21.77 $\pm$ 0.06 \\
    1999 Sep 30.13798 & 2451451.63798 & 21.82 $\pm$ 0.07 \\
    1999 Sep 30.14722 & 2451451.64722 & 21.74 $\pm$ 0.06 \\
    1999 Sep 30.15524 & 2451451.65524 & 21.72 $\pm$ 0.06 \\
    1999 Sep 30.16834 & 2451451.66834 & 21.72 $\pm$ 0.08 \\
    1999 Sep 30.17680 & 2451451.67680 & 21.83 $\pm$ 0.07 \\
    1999 Sep 30.18548 & 2451451.68548 & 21.80 $\pm$ 0.06 \\
    1999 Sep 30.19429 & 2451451.69429 & 21.74 $\pm$ 0.07 \\
    1999 Sep 30.20212 & 2451451.70212 & 21.78 $\pm$ 0.07 \\
    1999 Sep 30.21010 & 2451451.71010 & 21.72 $\pm$ 0.07 \\
    1999 Sep 30.21806 & 2451451.71806 & 21.76 $\pm$ 0.09 \\
    1999 Sep 30.23528 & 2451451.73528 & 21.73 $\pm$ 0.07 \\
    1999 Sep 30.24355 & 2451451.74355 & 21.74 $\pm$ 0.08 \\
    1999 Oct 01.02002 & 2451452.52002 & 21.81 $\pm$ 0.06 \\
    1999 Oct 01.02799 & 2451452.52799 & 21.82 $\pm$ 0.06 \\
    1999 Oct 01.03648 & 2451452.53648 & 21.81 $\pm$ 0.06 \\
    1999 Oct 01.04422 & 2451452.54422 & 21.80 $\pm$ 0.06 \\
    1999 Oct 01.93113 & 2451453.43113 & 21.71 $\pm$ 0.06 \\
    1999 Oct 01.94168 & 2451453.44168 & 21.68 $\pm$ 0.06 \\
    1999 Oct 01.95331 & 2451453.45331 & 21.73 $\pm$ 0.06 \\
    1999 Oct 01.97903 & 2451453.47903 & 21.69 $\pm$ 0.06 \\
    1999 Oct 01.99177 & 2451453.49177 & 21.74 $\pm$ 0.06 \\
    1999 Oct 02.00393 & 2451453.50393 & 21.73 $\pm$ 0.05 \\
    1999 Oct 02.01588 & 2451453.51588 & 21.78 $\pm$ 0.05 \\
    1999 Oct 02.02734 & 2451453.52734 & 21.71 $\pm$ 0.05 \\
    \enddata
    \tablenotetext{a}{UT date at the start of the exposure}
    \tablenotetext{b}{Julian date at the start of the exposure}
    \tablenotetext{c}{Apparent red magnitude; errors include
      uncertainties in relative and absolute photometry added
      quadratically}
  \end{deluxetable}

  \begin{deluxetable}{ccc}
    \tabletypesize{\scriptsize}
    \tablecaption{Photometric measurements of $\SN$
    \label{Table.1998SN165AbsPhot}}
    \tablewidth{0pt}
    \tablehead{
	\colhead{ } & \colhead{ } & \colhead{$m_R$\tnc} \\
	\colhead{UT Date\tna} & \colhead{Julian Date\tnb} & \colhead{[mag]}
    }
    \startdata
  1999 Sep 29.87384 & 2451451.37384 & 21.20 $\pm$ 0.06 \\
  1999 Sep 29.88050 & 2451451.38050 & 21.19 $\pm$ 0.05 \\
  1999 Sep 29.88845 & 2451451.38845 & 21.18 $\pm$ 0.05 \\
  1999 Sep 29.89811 & 2451451.39811 & 21.17 $\pm$ 0.05 \\
  1999 Sep 29.90496 & 2451451.40496 & 21.21 $\pm$ 0.05 \\
  1999 Sep 29.91060 & 2451451.41060 & 21.24 $\pm$ 0.05 \\
  1999 Sep 29.91608 & 2451451.41608 & 21.18 $\pm$ 0.05 \\
  1999 Sep 29.92439 & 2451451.42439 & 21.25 $\pm$ 0.05 \\
  1999 Sep 29.93055 & 2451451.43055 & 21.24 $\pm$ 0.05 \\
  1999 Sep 29.93712 & 2451451.43712 & 21.26 $\pm$ 0.06 \\
  1999 Sep 29.94283 & 2451451.44283 & 21.25 $\pm$ 0.06 \\
  1999 Sep 29.94821 & 2451451.44821 & 21.28 $\pm$ 0.06 \\
  1999 Sep 29.96009 & 2451451.46009 & 21.25 $\pm$ 0.06 \\
  1999 Sep 29.96640 & 2451451.46640 & 21.21 $\pm$ 0.06 \\
  1999 Sep 29.97313 & 2451451.47313 & 21.17 $\pm$ 0.05 \\
  1999 Sep 29.97850 & 2451451.47850 & 21.14 $\pm$ 0.05 \\
  1999 Sep 29.98373 & 2451451.48373 & 21.12 $\pm$ 0.06 \\
  1999 Sep 29.98897 & 2451451.48897 & 21.15 $\pm$ 0.06 \\
  1999 Sep 29.99469 & 2451451.49469 & 21.15 $\pm$ 0.06 \\
  1999 Sep 29.99997 & 2451451.49997 & 21.16 $\pm$ 0.06 \\
  1999 Sep 30.00521 & 2451451.50521 & 21.12 $\pm$ 0.06 \\
  1999 Sep 30.01144 & 2451451.51144 & 21.09 $\pm$ 0.06 \\
  1999 Sep 30.02164 & 2451451.52164 & 21.18 $\pm$ 0.06 \\
  1999 Sep 30.02692 & 2451451.52692 & 21.17 $\pm$ 0.06 \\
  1999 Sep 30.84539 & 2451452.34539 & 21.32 $\pm$ 0.06 \\
  1999 Sep 30.85033 & 2451452.35033 & 21.30 $\pm$ 0.06 \\
  1999 Sep 30.85531 & 2451452.35531 & 21.28 $\pm$ 0.06 \\
  1999 Sep 30.86029 & 2451452.36029 & 21.31 $\pm$ 0.06 \\
  1999 Sep 30.86550 & 2451452.36550 & 21.21 $\pm$ 0.06 \\
  1999 Sep 30.87098 & 2451452.37098 & 21.26 $\pm$ 0.06 \\
  1999 Sep 30.87627 & 2451452.37627 & 21.28 $\pm$ 0.06 \\
  1999 Sep 30.89202 & 2451452.39202 & 21.23 $\pm$ 0.05 \\
  1999 Sep 30.89698 & 2451452.39698 & 21.30 $\pm$ 0.06 \\
  1999 Sep 30.90608 & 2451452.40608 & 21.20 $\pm$ 0.05 \\
  1999 Sep 30.91191 & 2451452.41191 & 21.26 $\pm$ 0.05 \\
  1999 Sep 30.92029 & 2451452.42029 & 21.15 $\pm$ 0.05 \\
  1999 Sep 30.92601 & 2451452.42601 & 21.19 $\pm$ 0.05 \\
  1999 Sep 30.93110 & 2451452.43110 & 21.14 $\pm$ 0.05 \\
  1999 Sep 30.93627 & 2451452.43627 & 21.16 $\pm$ 0.05 \\
  1999 Sep 30.94858 & 2451452.44858 & 21.18 $\pm$ 0.05 \\
  1999 Sep 30.95363 & 2451452.45363 & 21.16 $\pm$ 0.05 \\
  1999 Sep 30.95852 & 2451452.45852 & 21.13 $\pm$ 0.05 \\
  1999 Sep 30.96347 & 2451452.46347 & 21.17 $\pm$ 0.05 \\
  1999 Sep 30.96850 & 2451452.46850 & 21.16 $\pm$ 0.05 \\
  1999 Sep 30.97422 & 2451452.47422 & 21.18 $\pm$ 0.05 \\
  1999 Sep 30.98431 & 2451452.48431 & 21.18 $\pm$ 0.05 \\
  1999 Sep 30.98923 & 2451452.48923 & 21.17 $\pm$ 0.06 \\
  1999 Sep 30.99444 & 2451452.49444 & 21.16 $\pm$ 0.05 \\
  1999 Sep 30.99934 & 2451452.49934 & 21.20 $\pm$ 0.05 \\
  1999 Oct 01.00424 & 2451452.50424 & 21.16 $\pm$ 0.05 \\
  1999 Oct 01.00992 & 2451452.50992 & 21.18 $\pm$ 0.06 \\
    \enddata
    \tablenotetext{a}{UT date at the start of the exposure}
    \tablenotetext{b}{Julian date at the start of the exposure}
    \tablenotetext{c}{Apparent red magnitude; errors include
      uncertainties in relative and absolute photometry added
      quadratically}
  \end{deluxetable}

  \begin{deluxetable}{ccc}
    \tabletypesize{\scriptsize}
    \tablecaption{Photometric measurements of (19521) Chaos
    \label{Table.1998WH24AbsPhot}}
    \tablewidth{0pt}
    \tablehead{
	\colhead{ } & \colhead{ } & \colhead{$m_R$\tnc} \\
	\colhead{UT Date\tna} & \colhead{Julian Date\tnb} & \colhead{[mag]}
    }
    \startdata
  1999 Oct 01.06329 & 2451452.56329 & 20.82 $\pm$ 0.06 \\
  1999 Oct 01.06831 & 2451452.56831 & 20.80 $\pm$ 0.06 \\
  1999 Oct 01.07324 & 2451452.57324 & 20.80 $\pm$ 0.06 \\
  1999 Oct 01.07817 & 2451452.57817 & 20.81 $\pm$ 0.06 \\
  1999 Oct 01.08311 & 2451452.58311 & 20.80 $\pm$ 0.06 \\
  1999 Oct 01.08801 & 2451452.58801 & 20.76 $\pm$ 0.06 \\
  1999 Oct 01.09370 & 2451452.59370 & 20.77 $\pm$ 0.06 \\
  1999 Oct 01.14333 & 2451452.64333 & 20.71 $\pm$ 0.06 \\
  1999 Oct 01.15073 & 2451452.65073 & 20.68 $\pm$ 0.06 \\
  1999 Oct 01.15755 & 2451452.65755 & 20.70 $\pm$ 0.06 \\
  1999 Oct 01.16543 & 2451452.66543 & 20.72 $\pm$ 0.06 \\
  1999 Oct 01.17316 & 2451452.67316 & 20.72 $\pm$ 0.06 \\
  1999 Oct 01.18112 & 2451452.68112 & 20.71 $\pm$ 0.06 \\
  1999 Oct 01.18882 & 2451452.68882 & 20.73 $\pm$ 0.06 \\
  1999 Oct 01.19652 & 2451452.69652 & 20.70 $\pm$ 0.06 \\
  1999 Oct 01.20436 & 2451452.70436 & 20.69 $\pm$ 0.06 \\
  1999 Oct 01.21326 & 2451452.71326 & 20.72 $\pm$ 0.06 \\
  1999 Oct 01.21865 & 2451452.71865 & 20.72 $\pm$ 0.06 \\
  1999 Oct 01.22402 & 2451452.72402 & 20.74 $\pm$ 0.06 \\
  1999 Oct 01.22938 & 2451452.72938 & 20.72 $\pm$ 0.06 \\
  1999 Oct 01.23478 & 2451452.73478 & 20.71 $\pm$ 0.07 \\
  1999 Oct 01.24022 & 2451452.74022 & 20.72 $\pm$ 0.07 \\
  1999 Oct 02.04310 & 2451453.54310 & 20.68 $\pm$ 0.06 \\
  1999 Oct 02.04942 & 2451453.54942 & 20.69 $\pm$ 0.06 \\
  1999 Oct 02.07568 & 2451453.57568 & 20.74 $\pm$ 0.07 \\
  1999 Oct 02.08266 & 2451453.58266 & 20.73 $\pm$ 0.06 \\
  1999 Oct 02.09188 & 2451453.59188 & 20.74 $\pm$ 0.06 \\
  1999 Oct 02.10484 & 2451453.60484 & 20.75 $\pm$ 0.06 \\
  1999 Oct 02.11386 & 2451453.61386 & 20.77 $\pm$ 0.06 \\
  1999 Oct 02.12215 & 2451453.62215 & 20.77 $\pm$ 0.06 \\
  1999 Oct 02.13063 & 2451453.63063 & 20.78 $\pm$ 0.06 \\
  1999 Oct 02.13982 & 2451453.63982 & 20.79 $\pm$ 0.06 \\
  1999 Oct 02.14929 & 2451453.64929 & 20.71 $\pm$ 0.07 \\
    \enddata
    \tablenotetext{a}{UT date at the start of the exposure}
    \tablenotetext{b}{Julian date at the start of the exposure}
    \tablenotetext{c}{Apparent red magnitude; errors include
      uncertainties in relative and absolute photometry added
      quadratically}
  \end{deluxetable}

  \begin{deluxetable}{ccr}
    \tabletypesize{\scriptsize}
    \tablecaption{Relative photometry measurements of $\DF$
    \label{Table.1999DF9RelPhot}}
    \tablewidth{0pt}
    \tablehead{
	\colhead{ } & \colhead{ } & \colhead{$\delta m_R$\tnc} \\
	\colhead{UT Date\tna} & \colhead{Julian Date\tnb} & \colhead{[mag]}
    }
    \startdata
2001 Feb 13.13417 & 2451953.63417 &  0.21 $\pm$ 0.02 \\
2001 Feb 13.14536 & 2451953.64536 &  0.20 $\pm$ 0.03 \\
2001 Feb 13.15720 & 2451953.65720 &  0.17 $\pm$ 0.04 \\
2001 Feb 13.16842 & 2451953.66842 &  0.06 $\pm$ 0.03 \\
2001 Feb 13.17967 & 2451953.67967 & -0.08 $\pm$ 0.02 \\
2001 Feb 13.20209 & 2451953.70209 & -0.09 $\pm$ 0.03 \\
2001 Feb 13.21325 & 2451953.71325 & -0.05 $\pm$ 0.03 \\
2001 Feb 13.22439 & 2451953.72439 & -0.15 $\pm$ 0.03 \\
2001 Feb 13.23554 & 2451953.73554 & -0.19 $\pm$ 0.04 \\
2001 Feb 13.24671 & 2451953.74671 &  0.00 $\pm$ 0.04 \\
2001 Feb 14.13972 & 2451954.63972 & -0.05 $\pm$ 0.02 \\
2001 Feb 14.15104 & 2451954.65104 & -0.12 $\pm$ 0.02 \\
2001 Feb 14.16228 & 2451954.66228 & -0.25 $\pm$ 0.02 \\
2001 Feb 14.17347 & 2451954.67347 & -0.18 $\pm$ 0.02 \\
2001 Feb 14.18477 & 2451954.68477 & -0.14 $\pm$ 0.03 \\
2001 Feb 14.19600 & 2451954.69600 & -0.05 $\pm$ 0.03 \\
2001 Feb 14.20725 & 2451954.70725 &  0.00 $\pm$ 0.03 \\
2001 Feb 14.21860 & 2451954.71860 &  0.03 $\pm$ 0.03 \\
2001 Feb 14.22987 & 2451954.72987 &  0.11 $\pm$ 0.04 \\
2001 Feb 14.24112 & 2451954.74112 &  0.21 $\pm$ 0.04 \\
2001 Feb 14.25234 & 2451954.75234 &  0.20 $\pm$ 0.05 \\
2001 Feb 14.26356 & 2451954.76356 &  0.16 $\pm$ 0.05 \\
2001 Feb 15.14518 & 2451955.64518 & -0.06 $\pm$ 0.05 \\
2001 Feb 15.15707 & 2451955.65707 & -0.08 $\pm$ 0.02 \\
2001 Feb 15.16831 & 2451955.66831 & -0.15 $\pm$ 0.05 \\
2001 Feb 15.19086 & 2451955.69086 &  0.05 $\pm$ 0.06 \\
2001 Feb 15.20234 & 2451955.70234 &  0.19 $\pm$ 0.07 \\
2001 Feb 15.23127 & 2451955.73127 &  0.04 $\pm$ 0.05 \\
    \enddata
    \tablenotetext{a}{UT date at the start of the exposure}
    \tablenotetext{b}{Julian date at the start of the exposure}
    \tablenotetext{c}{Mean-subtracted apparent red magnitude; errors include
      uncertainties in relative and absolute photometry added
      quadratically}
  \end{deluxetable}

  \begin{deluxetable}{ccr}
    \tabletypesize{\scriptsize}
    \tablecaption{Relative photometry measurements of $\CZ$
    \label{Table.2001CZ31RelPhot}}
    \tablewidth{0pt}
    \tablehead{
	\colhead{ } & \colhead{ } & \colhead{$\delta m_R$\tnc} \\
	\colhead{UT Date\tna} & \colhead{Julian Date\tnb} & \colhead{[mag]}
    }
    \startdata
2001 Feb 28.92789 & 2451969.42789 &  0.03 $\pm$ 0.05 \\
2001 Feb 28.93900 & 2451969.43900 &  0.06 $\pm$ 0.04 \\
2001 Feb 28.95013 & 2451969.45013 &  0.03 $\pm$ 0.04 \\
2001 Feb 28.96120 & 2451969.46120 & -0.09 $\pm$ 0.04 \\
2001 Feb 28.97235 & 2451969.47235 & -0.10 $\pm$ 0.04 \\
2001 Feb 28.98349 & 2451969.48349 & -0.12 $\pm$ 0.04 \\
2001 Feb 28.99475 & 2451969.49475 & -0.14 $\pm$ 0.03 \\
2001 Mar 01.00706 & 2451969.50706 & -0.02 $\pm$ 0.03 \\
2001 Mar 01.01817 & 2451969.51817 &  0.00 $\pm$ 0.03 \\
2001 Mar 01.02933 & 2451969.52933 &  0.03 $\pm$ 0.03 \\
2001 Mar 01.04046 & 2451969.54046 &  0.07 $\pm$ 0.04 \\
2001 Mar 01.05153 & 2451969.55153 &  0.10 $\pm$ 0.04 \\
2001 Mar 01.06304 & 2451969.56304 &  0.06 $\pm$ 0.04 \\
2001 Mar 01.08608 & 2451969.58608 & -0.05 $\pm$ 0.04 \\
2001 Mar 01.09808 & 2451969.59808 & -0.05 $\pm$ 0.05 \\
2001 Mar 03.01239 & 2451971.51239 &  0.15 $\pm$ 0.05 \\
2001 Mar 03.02455 & 2451971.52455 & -0.01 $\pm$ 0.05 \\
2001 Mar 03.03596 & 2451971.53596 &  0.00 $\pm$ 0.04 \\
2001 Mar 03.04731 & 2451971.54731 & -0.02 $\pm$ 0.03 \\
2001 Mar 03.05865 & 2451971.55865 & -0.08 $\pm$ 0.04 \\
2001 Mar 03.07060 & 2451971.57060 & -0.04 $\pm$ 0.04 \\
2001 Mar 03.08212 & 2451971.58212 &  0.01 $\pm$ 0.03 \\
    \enddata
    \tablenotetext{a}{UT date at the start of the exposure}
    \tablenotetext{b}{Julian date at the start of the exposure}
    \tablenotetext{c}{Mean-subtracted apparent red magnitude; errors include
      uncertainties in relative and absolute photometry added
      quadratically}
  \end{deluxetable}

\section{Lightcurve Analysis}

The results in this paper depend solely on the amplitude and period of
the KBO lightcurves. It is therefore important to accurately determine
these parameters and the associated uncertainties.

\subsection{Can we detect the KBO brightness variation?}
\label{Capitulo4KBOVariability}

We begin by investigating if the observed brightness variations are
intrinsic to the KBO, i.e., if the KBO's intrinsic brightness
variations are detectable given our uncertainties. This was done by
comparing the frame-to-frame scatter in the KBO optical data with that
of ($\sim10-20$) comparison stars.

To visually compare the scatter in the magnitudes of the reference
stars (see Section \ref{Capitulo4Sec.Observations}), comparison stars,
and KBOs, we plot a histogram of their frame-to-frame variances (see
Fig.~\ref{Fig.VarHistograms}). In general such a histogram should show
the reference stars clustered at the lowest variances, followed by the
comparison stars spread over larger variances. If the KBO appears
isolated at much higher variances than both groups of stars (e.g.,
Fig.~\ref{Fig.VarHistograms}j), then its brightness modulations are
significant. Conversely, if the KBO is clustered with the stars (e.g.
Fig.~\ref{Fig.VarHistograms}f), any periodic brightness variations
would be below the detection threshold.

Figure~\ref{Fig.VarvsMag} shows the dependence of the uncertainties on
magnitude. Objects that do not fall on the rising curve traced out by
the stars must have intrinsic brightness variations. By calculating
the mean and spread of the variance for the comparison stars (shown as
crosses) we can calculate our photometric uncertainties and thus
determine whether the KBO brightness variations are significant
($\ge$3$\sigma$).

\subsection{Period determination}
\label{Capitulo4Period.Determination}

In the cases where significant brightness variations (see
Section~\ref{Capitulo4KBOVariability}) were detected in the
lightcurves, the phase dispersion minimization method was used
\citep[PDM,][]{1978ApJ...224..953S} to look for periodicities in the
data. For each test period, PDM computes a statistical parameter
$\theta$ that compares the spread of data points in phase bins with
the overall spread of the data. The period that best fits the data is
the one that minimizes $\theta$. The advantages of PDM are that it is
non-parametric, i.e., it does not assume a shape for the lightcurve,
and it can handle unevenly spaced data.  Each data set was tested for
periods ranging from 2 to 24 hours, in steps of 0.01$\,$hr.
To assess the uniqueness of the PDM solution, we generated 100 Monte 
Carlo realizations of each lightcurve, keeping the original data times 
and randomizing the magnitudes within the error bars. We ran PDM on each 
generated dataset to obtain a distribution of best-fit periods.

  \begin{deluxetable*}{r@{ }lcccc}
    \tabletypesize{\scriptsize}
    \tablecaption{Lightcurve Properties of Observed KBOs
    \label{Table.LCProperties}}
    \tablewidth{0pt}
    \tablehead{
	\multicolumn{2}{c}{Object Designation} & \colhead{$m_R$\tna} &
	\colhead{Nts\tnb} & \colhead{$\Delta m_R$\tnc} & 
	\colhead{$P$\tnd} \\
	\multicolumn{2}{c}{ } & \colhead{[mag]} & \colhead{ } &
	\colhead{[mag]} & \colhead{[hr]}
    }
    \startdata
  (35671) &$\SNnn$&  21.20$\pm$0.05 & 2(1) & 0.16$\pm$0.01 & 8.84 (8.70) \\
  (79983) &$\DFnn$& --              & 3    & 0.40$\pm$0.02 & 6.65 (9.00)\\
          & $\CZ$ &                 & 2(1) & 0.21$\pm$0.02 & 4.71 (5.23)\\
          &       &                 &      &               &      \\
  (19308) &$\TOnn$& 21.26$\pm$0.06  & 1    & ?             &      \\
          & $\TS$ & 21.76$\pm$0.05  & 3    & $<$0.15       &      \\ 
  (19521) & Chaos & 20.74$\pm$0.06  & 2    & $<$0.10       &      \\
  (80806) &$\CMnn$& --              & 2    & $<$0.14       &      \\
  (66652) &$\RZnn$& --              & 3    & $<$0.05       &      \\
  (47171) &$\TCnn$& --              & 3    & $<$0.07       &      \\
  (38628) & Huya  & --              & 2    & $<$0.04       &      \\
    \enddata
    \tablenotetext{a}{nean red magnitude. Errors include uncertainties
      in relative and absolute photometry added quadratically;}
    \tablenotetext{b}{number of nights with useful data. Numbers in
      brackets indicate number of nights of data from other observers
      used for period determination. Data for $\SN$ taken from
      \citet{2002NewA....7..359P} and data for $\CZ$ taken from SJ02;}
    \tablenotetext{c}{lightcurve amplitude;}
    \tablenotetext{d}{lightcurve period (values in parenthesis indicate 
      less likely solutions not entirely ruled out by the data).}
  \end{deluxetable*}

\subsection{Amplitude determination}
\label{Capitulo4Amplitude.Determination}

We used a Monte Carlo experiment to determine the amplitude of the
lightcurves for which a period was found. We generated several
artificial data sets by randomizing each point within the error
bar. Each artificial data set was fitted with a Fourier series, using
the best-fit period, and the mode and central 68\% of the distribution
of amplitudes were taken as the lightcurve amplitude and $1\sigma$
uncertainty, respectively.

For the null lightcurves, i.e. those where no significant variation
was detected, we subtracted the typical error bar size from the total
amplitude of the data to obtain an upper limit to the amplitude of the
KBO photometric variation.

\section{Results}

In this section we present the results of the lightcurve analysis for
each of the observed KBOs. We found significant brightness variations
($\Delta m>0.15\,$mag) for 3 out of 10 KBOs (30\%), and $\Delta m \ge
0.40\,$mag for 1 out of 10 (10\%). This is consistent with previously
published results: \citeauthor{2002AJ....124.1757S}
(\citeyear{2002AJ....124.1757S}, hereafter
\citetalias{2002AJ....124.1757S}) found a fraction of 31\% with
$\Delta m>0.15\,$mag and 23\% with $\Delta m>0.40\,$mag, both
consistent with our results.  The other 7 KBOs do not show detectable
variations. The results are summarized in
Table~\ref{Table.LCProperties}.

\subsection{1998$\,$SN$_\mathbf{165}$}

\FigC

\FigD

The brightness of $\SN$ varies significantly ($>5\sigma$) more than
that of the comparison stars (see Figs.~\ref{Fig.VarvsMag} and
\ref{Fig.VarHistograms}c).  The periodogram for this KBO shows a very
broad minimum around $P=9\,$hr
(Fig.~\ref{Fig.1998SN165Periodogram}a). The degeneracy implied by the
broad minimum would only be resolved with additional data. A slight
weaker minimum is seen at $P=6.5\,$hr, which is close to a 24$\,$hr
alias of $P=9\,$hr.

\citeauthor{2002NewA....7..359P} (\citeyear{2002NewA....7..359P},
hereafter \citetalias{2002NewA....7..359P}) observed this object in
September 2000, but having only one night's worth of data, they did
not succeed in determining this object's rotational period
unambiguously. We used their data to solve the degeneracy in our PDM
result. The \citetalias{2002NewA....7..359P} data have not been
absolutely calibrated, and the magnitudes are given relative to a
bright field star. To be able to combine it with our own data we had
to subtract the mean magnitudes. Our periodogram of $\SN$ (centered on
the broad minimum) is shown in Fig.~\ref{Fig.1998SN165Periodogram}b
and can be compared with the revised periodogram obtained with our
data combined with the \citetalias{2002NewA....7..359P} data
(Fig.~\ref{Fig.1998SN165Periodogram}c).  The minima become much
clearer with the additional data, but because of the 1-year time
difference between the two observational campaigns, many close aliases
appear in the periodogram.  The absolute minimum, at $P=8.84\,$hr,
corresponds to a double peaked lightcurve (see
Fig.~\ref{Fig.1998SN1650884fit}). The second best fit, $P=8.7\,$hr,
produces a more scattered phase plot, in which the peak in the
\citetalias{2002NewA....7..359P} data coincides with our night 2. 
Period $P=8.84\,$hr was also favored by the Monte Carlo method
described in Section~\ref{Capitulo4Period.Determination}, being identified
as the best fit in 55\% of the cases versus 35\% for $P=8.7\,$hr. The large
size of the error bars compared to the amplitude is responsible for 
the ambiguity in the result. We will use $P=8.84\,$hr in the rest of 
the paper because it was consistently selected as the best fit.

The amplitude, obtained using the Monte Carlo method described in
Section~\ref{Capitulo4Amplitude.Determination}, is $\Delta
m=0.16\pm0.01\,$mag. This value was calculated using only our data,
but it did not change when recalculated adding the
\citetalias{2002NewA....7..359P} data.

\subsection{1999$\,$DF$_\mathbf{9}$}

\FigE

\FigF

$\DF$ shows large amplitude photometric variations ($\Delta
m_R\sim0.4\,$mag). The PDM periodogram for $\DF$ is shown in
Fig.~\ref{Fig.1999DF9.periodogram}. The best-fit period is
$P=6.65\,$hr, which corresponds to a double-peak lightcurve
(Fig.~\ref{Fig.1999DF90665fit}).  Other PDM minima are found close to
$P/2\approx3.3\,$hr and $9.2\,$hr, a $24\,$hr alias of the best
period. Phasing the data with $P/2$ results in a worse fit because the
two minima of the double peaked lightcurve exhibit significantly
different morphologies (Fig.~\ref{Fig.1999DF90665fit}); the peculiar sharp
feature superimposed on the brighter minimum, which is reproduced on two different
nights, may be caused by a non-convex feature on the surface of the KBO
\citep{2003Icar..164..346T}. Period $P=6.65\,$hr was selected in
65 of the 100 Monte Carlo replications of the dataset
(see Section~\ref{Capitulo4Period.Determination}). The second most selected
solution (15\%) was at $P=9\,$hr. We will use $P=6.65\,$hr for the rest of
the paper. 

The amplitude of the lightcurve, estimated as described in
Section~\ref{Capitulo4Amplitude.Determination}, is $\Delta
m_R=0.40\pm0.02\,$mag.

\subsection{2001$\,$CZ$_\mathbf{31}$}

\FigG

\FigH

This object shows substantial brightness variations ($4.5\sigma$ above
the comparison stars) in a systematic manner. The first night of data
seems to sample nearly one complete rotational phase. As for $\SN$,
the $\CZ$ data span only two nights of observations.  In this case,
however, the PDM minima (see Figs.~\ref{Fig.2001CZ31.periodogram}a and
\ref{Fig.2001CZ31.periodogram}b) are very narrow and only two
correspond to independent periods, $P=4.69\,$hr (the minimum at
$5.82\,$ is a $24\,$hr alias of $4.69\,$hr), and $P=5.23\,$hr.

$\CZ$ has also been observed by \citetalias{2002AJ....124.1757S} in
February and April 2001, with inconclusive results. We used their data
to try to rule out one (or both) of the two periods we found. We
mean-subtracted the \citetalias{2002AJ....124.1757S} data in
order to combine it with our uncalibrated
observations. Figure~\ref{Fig.2001CZ31.periodogram}c shows the section
of the periodogram around $P=5\,$hr, recalculated using
\citetalias{2002AJ....124.1757S}'s first night plus our own data. The
aliases are due to the 1 month time difference between the two
observing runs. The new PDM minimum is at $P=4.71\,$hr -- very close
to the $P=4.69\,$hr determined from our data alone.

Visual inspection of the combined data set phased with $P=4.71\,$hr
shows a very good match between \citetalias{2002AJ....124.1757S}'s
first night (2001~Feb~20) and our own data (see
Fig.~\ref{Fig.2001CZ31.0471fit}). \citetalias{2002AJ....124.1757S}'s
second and third nights show very large scatter and were not included
in our analysis.  Phasing the data with $P=5.23\,$hr yields a more
scattered lightcurve, which confirms the PDM result.  The Monte Carlo test
for uniqueness yielded $P=4.71\,$hr as the best-fit period in 57\% of the 
cases, followed by $P=5.23\,$hr in 21\%, and a few other solutions, all below 
10\%, between $P=5\,$hr and $P=6\,$hr. We will use $P=4.71\,$hr throughout 
the rest of the paper.

We measured a lightcurve amplitude of $\Delta m=0.21 \pm
0.02\,$mag. If we use both ours and \citetalias{2002AJ....124.1757S}'s
first night data, $\Delta m$ rises to 0.22$\,$mag.

\subsection{Flat Lightcurves}

\FigI

The fluctuations detected in the optical data on KBOs $\TO$, $\TS$,
$\TC$, $\RZ$, $\CM$, and $\EB$, are well within the
uncertainties. $\WH$ shows some variations with respect to the
comparison stars but no period was found to fit all the data. See
Table~\ref{Table.LCProperties} and Fig.~\ref{Fig.FlatLCurves} for a
summary of the results.

\subsection{Other lightcurve measurements}

The KBO lightcurve database has increased considerably in the last few
years, largely due to the observational campaign of
\citetalias{2002AJ....124.1757S}, with recent updates in
\citet{2003EM&P...92..207S} and \citet{2004AJ....127.3023S}. These
authors have published observations and rotational data for a total of
30 KBOs (their \citetalias{2002AJ....124.1757S} paper includes data
for 3 other previously published lightcurves in the analysis). Other
recently published KBO lightcurves include those for $(50000)\,$Quaoar
\citep{2003A&A...409L..13O} and the scattered KBO
$(29981)\,1999\,\rm{TD}_{10}$ \citep{2003A&A...407.1139R}. Of the 10
KBO lightcurves presented in this paper, 6 are new to the database,
bringing the total to 41.

Table~\ref{Table.OtherLC} lists the rotational data on the 41 KBOs
that will be analyzed in the rest of the paper.

  \begin{deluxetable*}{r@{ }lcrr@{}lcc}
    \tabletypesize{\scriptsize}
    \tablecaption{Database of KBOs Lightcurve Properties
    \label{Table.OtherLC}}
    \tablewidth{0pt}
    \tablehead{
	\multicolumn{2}{c}{Object Designation} & \colhead{Class\tna} &
	\colhead{Size\tnb} & \multicolumn{2}{c}{$P$\tnc} & 
	\colhead{$\Delta m_R$\tnd} & \colhead{Ref.} \\
	\multicolumn{3}{c}{ } & \colhead{[km]} & \multicolumn{2}{c}{[hr]} &
	\colhead{[mag]} & \colhead{ }
    }
    \startdata
    \cutinhead{KBOs considered to have $\Delta m < 0.15\,$mag}
(15789) & $1993\,\rm{SC} $      & P  & 240 & &  &    0.04       & 7, 2       \\
(15820) & $1994\,\rm{TB}$       & P  & 220 & &  & $<$0.04       & 10         \\
(26181) & $1996\,\rm{GQ}_{21}$  & S  & 730 & &  & $<$0.10       & 10         \\
(15874) & $1996\,\rm{TL}_{66}$  & S  & 480 & &  &    0.06       & 7, 4       \\
(15875) & $1996\,\rm{TP}_{66}$  & P  & 250 & &  &    0.12       & 7, 1       \\
(79360) & $1997\,\rm{CS}_{29}$  & C  & 630 & &  & $<$0.08       & 10         \\
(91133) & $1998\,\rm{HK}_{151}$ & P  & 170 & &  & $<$0.15       & 10         \\
(33340) & $1998\,\rm{VG}_{44}$  & P  & 280 & &  & $<$0.10       & 10         \\
(19521) & Chaos                 & C  & 600 & &  & $<$0.10       & 13, 10     \\
(26375) & $1999\,\rm{DE}_{9}$   & S  & 700 & &  & $<$0.10       & 10         \\
(47171) & $1999\,\rm{TC}_{36}$  & Pb & 300 & &  & $<$0.07       & 13, 11     \\
(38628) & Huya                  & P  & 550 & &  & $<$0.04       & 13, 11     \\
(82075) & $2000\,\rm{YW}_{134}$ & S  & 790 & &  & $<$0.10       & 11         \\
(82158) & $2001\,\rm{FP}_{185}$ & S  & 400 & &  & $<$0.10       & 11         \\
(82155) & $2001\,\rm{FZ}_{173}$ & S  & 430 & &  & $<$0.06       & 10         \\
        & $2001\,\rm{KD}_{77}$  & P  & 430 & &  & $<$0.10       & 11         \\
(28978) & Ixion                 & P  & 820 & &  & $<$0.10       & 11, 5      \\
        & $2001\,\rm{QF}_{298}$ & P  & 580 & &  & $<$0.10       & 11         \\
(42301) & $2001\,\rm{UR}_{163}$ & S  &1020 & &  & $<$0.10       & 11         \\
(42355) & $2002\,\rm{CR}_{46}$  & S  & 210 & &  & $<$0.10       & 11         \\
(55636) & $2002\,\rm{TX}_{300}$ & C  & 710 & 16.&24      & 0.08$\pm$0.02 & 11\\
(55637) & $2002\,\rm{UX}_{25}$  & C  &1090 & &  & $<$0.10       & 11         \\
(55638) & $2002\,\rm{VE}_{95}$  & P  & 500 & &  & $<$0.10       & 11         \\
(80806) & $2000\,\rm{CM}_{105}$ & C  & 330 & &  & $<$0.14       & 13         \\
(66652) & $1999\,\rm{RZ}_{253}$ & Cb & 170 & &  & $<$0.05       & 13         \\
        & $1996\,\rm{TS}_{66}$  & C  & 300 & &  & $<$0.14       & 13         \\
    \cutinhead{KBOs considered to have $\Delta m \geq 0.15\,$mag}
(32929) & $1995\,\rm{QY}_{9}$   & P  & 180 & 7.&3   & 0.60$\pm$0.04 & 10, 7  \\
(24835) & $1995\,\rm{SM}_{55}$  & C  & 630 & 8.&08  & 0.19$\pm$0.05 & 11     \\
(19308) & $1996\,\rm{TO}_{66}$  & C  & 720 & 7.&9   & 0.26$\pm$0.03 & 11, 3  \\
(26308) & $1998\,\rm{SM}_{165}$ & R  & 240 & 7.&1   & 0.45$\pm$0.03 & 10, 8  \\
(33128) & $1998\,\rm{BU}_{48}$  & S  & 210 & 9.&8   & 0.68$\pm$0.04 & 10, 8  \\
(40314) & $1999\,\rm{KR}_{16}$  & C  & 400 &11.&858 & 0.18$\pm$0.04 & 10     \\
(47932) & $2000\,\rm{GN}_{171}$ & P  & 360 & 8.&329 & 0.61$\pm$0.03 & 10     \\
(20000) & Varuna                & C  & 980 & 6.&34  & 0.42$\pm$0.03 & 10     \\
        & $2003\,\rm{AZ}_{84}$  & P  & 900 &13.&44  & 0.14$\pm$0.03 & 11     \\
        & $2001\,\rm{QG}_{298}$ & Pcb& 240 &13.&7744& 1.14$\pm$0.04 & 12     \\
(50000) & Quaoar                & C  &1300 &17.&6788& 0.17$\pm$0.02 & 6      \\
(29981) & $1999\,\rm{TD}_{10}$  & S  & 100 &15.&3833& 0.53$\pm$0.03 & 9      \\
(35671) & $1998\,\rm{SN}_{165}$ & C  & 400 & 8.&84  & 0.16$\pm$0.01 & 13     \\
(79983) & $1999\,\rm{DF}_{9}$   & C  & 340 & 6.&65  & 0.40$\pm$0.02 & 13     \\
        & $2001\,\rm{CZ}_{31}$  & C  & 440 & 4.&71  & 0.21$\pm$0.06 & 13     \\
    \enddata
    \tablenotetext{a}{Dynamical class (C = classical KBO, P = Plutino,
      R = 2:1 Resonant b = binary KBO);}
    \tablenotetext{b}{Diameter in km assuming an albedo of 0.04 except 
      when measured (see text);}
    \tablenotetext{c}{Period of the lightcurve in hours. For KBOs with
      both single and double peaked possible lightcurves the double
      peaked period is listed;}
    \tablenotetext{d}{\mbox{Lightcurve} amplitude in magnitudes.}
    \tablerefs{
      (1) \citet{1999MNRAS.308..588C}; (2) \citet{1997Icar..125...61D}; 
      (3) \citet{2000A&A...356.1076H}; (4) \citet{1998ApJ...494L.117L};
      (5) \citet{2001DPS....33.1211O}; (6) \citet{2003A&A...409L..13O};
      (7) \citet{1999Natur.398..129R}; (8) \citet{2001DPS....33.0606R};
      (9) \citet{2003A&A...407.1139R}; (10) \citet{2002AJ....124.1757S};
      (11) \citet{2003EM&P...92..207S}; (12) \citet{2004AJ....127.3023S}; 
      (13) this work.}
  \end{deluxetable*}

\section{Analysis}

In this section we examine the lightcurve properties of KBOs and
compare them with those of main-belt asteroids (MBAs).  The lightcurve
data for these two families of objects cover different size
ranges. MBAs, being closer to Earth, can be observed down to much
smaller sizes than KBOs; in general it is very difficult to obtain
good quality lightcurves for KBOs with diameters
$D<50\,$km. Furthermore, some KBOs surpass the $1000\,$km barrier
whereas the largest asteroid, Ceres, does not reach $900\,$km. This
will be taken into account in the analysis.

The lightcurve data for asteroids were taken from the Harris
Lightcurve Catalog\footnote{
  \url{http://pdssbn.astro.umd.edu/sbnhtml/asteroids/colors\_lc.html}
}, Version 5, while the diameter data were obtained from the Lowell
Observatory database of asteroids orbital elements\footnote{
  \url{ftp://ftp.lowell.edu/pub/elgb/astorb.html} }.  The sizes of
most KBOs were calculated from their absolute magnitude assuming an
albedo of 0.04. The exceptions are (47171)$\,$1999$\,$TC$_{36}$,
(38638)$\,$Huya, (28978)$\,$Ixion, (55636)$\,$2002$\,$TX$_{36}$,
(66652)$\,$1999$\,$RZ$_{36}$, (26308)$\,$1998$\,$SM$_{165}$, and
(20000)$\,$Varuna for which the albedo has been shown to be
inconsistent with the value 0.04 \citep{2005Icar..176..184G}. For
example, in the case of (20000)$\,$Varuna simultaneous thermal and
optical observations have yielded a red geometric albedo of
0.070$_{-0.017}^{+0.030}$ \citep{2001Natur.411..446J}.

\subsection{Spin period statistics}

\FigJ

As Fig.~\ref{Fig.PeriodDistrib} shows, the spin period distributions
of KBOs and MBAs are significantly different. Because the sample of
KBOs of small size or large periods is poor, to avoid bias in our
comparison we consider only KBOs and MBAs with diameter larger than
$200\,$km and with periods below $20\,$hr. In this range the mean
rotational periods of KBOs and MBAs are $9.23\,$hr and $6.48\,$hr,
respectively, and the two are different with a 98.5\% confidence
according to Student's $t$-test. However, the different means do not
rule that the underlying distributions are the same, and could simply
mean that the two sets of data sample the same distribution
differently. This is not the case, however, according to the
Kolmogorov-Smirnov (K-S) test, which gives a probability that the
periods of KBOs and MBAs are drawn from the same parent distribution
of 0.7\%.

Although it is clear that KBOs spin slower than asteroids, it is not clear
why this is so.  If collisions have contributed as significantly to the angular
momentum of KBOs as they did for MBAs
\citep{1982Icar...52..409F,1984A&A...138..464C}, then the observed difference
should be related to how these two families react to collision events.  We will
address the question of the collisional evolution of KBO spin rates in a future
paper.

\subsection{Lightcurve amplitudes and the shapes of KBOs}
\label{Capitulo4Shape.Distribution}

\FigK

\FigL

\FigM

The cumulative distribution of KBO lightcurve amplitudes is shown in
Fig.~\ref{Fig.CumulAmpl}.  It rises very steeply in the low amplitude range
($\Delta m < 0.15\,$mag), and then becomes shallower reaching large amplitudes.
In quantitative terms, $\sim 70\%$ of the KBOs possess $\Delta m < 0.15\,$mag,
while $\sim 12\%$ possess $\Delta m \ge 0.40\,$mag, with the maximum value
being $\Delta m=0.68\,$mag.  [Note: Fig.~\ref{Fig.CumulAmpl} does not include
the KBO 2001$\,$QG$_{298}$ which has a lightcurve amplitude $\Delta
m=1.14\pm0.04\,$mag, and would further extend the range of amplitudes.  We do
not include 2001$\,$QG$_{298}$ in our analysis because it is thought to be a
contact binary \citep{2004AJ....127.3023S}]. Figure~\ref{Fig.CumulAmpl} also
compares the KBO distribution with that of MBAs.  The distributions of the two
populations are clearly distinct: there is a larger fraction of KBOs in the low
amplitude range ($\Delta m < 0.15\,$mag) than in the case of MBAs, and the KBO
distribution extends to larger values of $\Delta m$.

Figure~\ref{Fig.AmplvsSize} shows the lightcurve amplitude of KBOs and MBAs
plotted against size.  KBOs with diameters larger than $D=400\,$km seem to have
lower lightcurve amplitudes than KBOs with diameters smaller than $D=400\,$km.
Student's $t$-test confirms that the mean amplitudes in each of these two size
ranges are different at the 98.5\% confidence level. For MBAs the transition is
less sharp and  seems to occur at a smaller size ($D\sim 200\,$km). In the case
of asteroids, the accepted explanation is that small bodies
($D\lesssim100\,$km) are fragments of high-velocity impacts, whereas of their
larger counterparts ($D>200\,$km) generally are not
\citep{1984A&A...138..464C}.  The lightcurve data on small KBOs are still too
sparse to permit a similar analysis.  In order to reduce the effects of bias
related to body size, we can consider only those KBOs and MBAs with diameters
larger than 200$\,$km.  In this size range, 25 of 37 KBOs (69\%) and 10 of 27
MBAs (37\%) have lightcurve amplitudes below 0.15$\,$mag. We used the Fisher
exact test to calculate the probability that such a contingency table would
arise if the lightcurve amplitude distributions of KBOs and MBAs were the same:
the resulting probability is 0.8\%. 

The distribution of lightcurve amplitudes can be used to infer the shapes of
KBOs, if certain reasonable assumptions are made (see, e.g.,
\citetalias{2003Icar..161..174L}).  Generally, objects with elongated shapes
produce large brightness variations due to their changing projected
cross-section as they rotate. Conversely, round objects, or those with the spin
axis aligned with the line of sight, produce little or no brightness
variations, resulting in "flat" lightcurves.  Figure~\ref{Fig.AmplvsSize} shows
that the lightcurve amplitudes of KBOs with diameters smaller and larger than
$D=400\,$km are significantly different.  Does this mean that the shapes of
KBOs are also different in these two size ranges? To investigate this
possibility of a size dependence among KBO shapes we will consider KBOs with
diameter smaller and larger than $400\,$km separately. We shall loosely refer
to objects with diameter $D>400\,$km and $D \le 400\,$km as {\em larger} and
{\em smaller} KBOs, respectively.

We approximate the shapes of KBOs by triaxial ellipsoids with semi-axes
$a>b>c$. For simplicity we consider the case where $b=c$ and use the axis ratio
$\tilde{a}=a/b$ to characterize the shape of an object. The orientation of the
spin axis is parameterized by the aspect angle $\theta$, defined as the
smallest angular distance between the line of sight and the spin vector. On
this basis the lightcurve amplitude $\Delta m$ is related to $\tilde{a}$ and
$\theta$ via the relation (Eq. (2) of \citetalias{2003Icar..161..174L} with
$\bar{c}=1$)
   \begin{eqnarray}
     \Delta m=2.5\log \sqrt{ \frac{2\,\tilde{a}^2} {1 + \tilde{a}^2 +
       \left( \tilde{a}^2 - 1 \right) \,\cos (2\,\theta )}}\;.
     \label{EqnDeltaMagCap4}
   \end{eqnarray}

\noindent Following \citetalias{2003EM&P...92..221L} we model the shape
distribution by a power-law of the form
   \begin{eqnarray}
     f(\tilde{a})\,{\rm d}\tilde{a}\propto\tilde{a}^{-q}\,{\rm
     d}\tilde{a}
     \label{EqnShapeDist}
   \end{eqnarray}
\noindent where $f(\tilde{a})\,{\rm d}\tilde{a}$ represents the fraction of
objects with shapes between $\tilde{a}$ and $\tilde{a}+{\rm d}\tilde{a}$. We
use the measured lightcurve amplitudes to estimate the value of $q$ by
employing both the method described in \citetalias{2003Icar..161..174L}, and by
Monte Carlo fitting the observed amplitude distribution
\citepalias{2002AJ....124.1757S,2003EM&P...92..221L}.  The latter consists of
generating artificial distributions of $\Delta m$ (Eq.~\ref{EqnDeltaMagCap4})
with values of $\tilde{a}$ drawn from distributions characterized by different
$q$'s (Eq.~\ref{EqnShapeDist}), and selecting the one that best fits the
observed cumulative amplitude distribution (Fig.~\ref{Fig.CumulAmpl}).  The
values of $\theta$ are generated assuming random spin axis orientations.  We
use the K-S test to compare the different fits. The errors are derived by
bootstrap resampling the original data set \citep{Efron79}, and measuring the
dispersion in the distribution of best-fit power-law indexes, $q_i$, found for
each bootstrap replication.

Following the \citetalias{2003Icar..161..174L} method we calculate the
probability of finding a KBO with $\Delta m \ge 0.15\,$mag:
   \begin{eqnarray}
     p(\Delta m \ge 0.15) \approx \int^{\tilde{a}_{\rm
        max}}_{\sqrt{K}} f(\tilde{a})\,\sqrt{\frac{\tilde{a}^2 -
        K}{(\tilde{a}^2 - 1) K}} \,\mathrm{d}\tilde{a}.
     \label{EqnProb2}
   \end{eqnarray}   
\noindent where $K=10^{0.8\times0.15}$, $f(\tilde{a})=C\,\tilde{a}^{-q}$, and
$C$ is a normalization constant.  This probability is calculated for a range of
$q$'s to determine the one that best matches the observed fraction of
lightcurves with amplitude larger than 0.15$\,$mag. These fractions are
$f(\Delta m\ge0.15\,{\rm mag};\,D\le400\,{\rm km})=8/19$, and $f(\Delta
m\ge0.15\,{\rm mag};\,D>400\,{\rm km})=5/21$, and $f(\Delta m\ge0.15\,{\rm
mag})=13/40$ for the complete set of data.  The results are summarized in
Table~\ref{Table.BestFitShapeDist} and shown in Fig.~\ref{Fig.CumulDistrFit}.
   
The uncertainties in the values of $q$ obtained using the
\citetalias{2003Icar..161..174L} method ($q=4.3^{+2.0}_{-1.6}$ for KBOs with
$D\le400\,$km and $q=7.4^{+3.1}_{-2.4}$ for KBOs with $D>400\,$km ; see
Table~\ref{Table.BestFitShapeDist}) do not rule out similar shape distributions
for smaller and larger KBOs. This is not the case for the Monte Carlo method.
The reason for this is that the \citetalias{2003Icar..161..174L} method relies
on a single, more robust parameter: the fraction of lightcurves with detectable
variations. The sizeable error bar is indicative that a larger dataset is
needed to better constrain the values of $q$.  In any case, it is reassuring
that both methods yield steeper shape distributions for larger KBOs, implying
more spherical shapes in this size range. A distribution with $q\sim8$ predicts
that $\sim$75\% of the large KBOs have $a/b<1.2$. For the smaller objects we
find a shallower distribution, $q\sim4$, which implies a significant fraction
of very elongated objects: $\sim$20\% have $a/b>1.7$.  Although based on small
numbers, the shape distribution of large KBOs is well fit by a simple power-law
(the K-S rejection probability is 0.6\%). This is not the case for smaller KBOs
for which the fit is poorer (K-S rejection probability is 20\%, see
Fig.~\ref{Fig.CumulDistrFit}).  Our results are in agreement with previous
studies of the overall KBO shape distribution, which had already shown that a
simple power-law does not explain the shapes of KBOs as a whole
\citepalias{2003EM&P...92..221L,2002AJ....124.1757S}.

The results presented in this section suggest that the shape distributions of
smaller and larger KBOs are different. However, the existing number of
lightcurves is not enough to make this difference statistically significant.
When compared to asteroids, KBOs show a preponderance of low amplitude
lightcurves, possibly a consequence of their possessing a larger fraction of
nearly spherical objects.  It should be noted that most of our analysis assumes
that the lightcurve sample used is homogeneous and unbiased; this is probably
not true. Different observing conditions, instrumentation, and data analysis
methods introduce systematic uncertainties in the dataset.  However, the most
likely source of bias in the sample is that some flat lightcurves may not have
been published. If this is the case, our conclusion that the amplitude
distributions of KBOs and MBAs are different would be strengthened. On the
other hand, if most unreported non-detections correspond to smaller KBOs then
the inferred contrast in the shape distributions of different-sized KBOs would
be less significant. Clearly, better observational contraints, particularly of
smaller KBOs, are necessary to constrain the KBO shape distribution and
understand its origin. 

  \begin{deluxetable}{ccc}
    \tabletypesize{\scriptsize}
    \tablecaption{Best fit parameter to the KBO shape distribution
    \label{Table.BestFitShapeDist}}
    \tablewidth{0pt}
    \tablehead{
	\colhead{ } & \multicolumn{2}{c}{Method\tnb} \\
	\colhead{Size Range\tna} & \colhead{LL03} & \colhead{MC}
    }
    \startdata
    $D \le 400\,$km & $q=4.3_{-1.6}^{+2.0}$  & $q=3.8\pm0.8$\\
    $D  >  400\,$km & $q=7.4_{-2.4}^{+3.1}$  & $q=8.0\pm1.4$\\
    All sizes       & $q=5.7_{-1.3}^{+1.6}$  & $q=5.3\pm0.8$\\
    \enddata
    \tablenotetext{a}{Range of KBO diameters, in km, considered in each
     case;}
    \tablenotetext{b}{\mbox{LL03} is the method described in
     \citet{2003Icar..161..174L}, and MC is a Monte Carlo fit of the
     lightcurve amplitude distribution.}
  \end{deluxetable}

\subsection{The inner structure of KBOs}

\FigN

In this section we wish to investigate if the rotational properties of KBOs
show any evidence that they have a rubble pile structure; a possible dependence
on object size is also investigated.  As in the case of asteroids, collisional
evolution may have played an important role in modifying the inner structure of
KBOs. Large asteroids ($D\gtrsim200\,$km) have in principle survived
collisional destruction for the age of the solar system, but may nonetheless
have been converted to rubble piles by repeated impacts. As a result of
multiple collisions, the ``loose'' pieces of the larger asteroids may have
reassembled into shapes close to triaxial equilibrium ellipsoids
\citep{1981Icar...46..114F}. Instead, the shapes of smaller asteroids
($D\le100\,$km) are consistent with collisional fragments
\citep{1984A&A...138..464C}, indicating that they are most likely by-products
of disruptive collisions.

Figure~\ref{Fig.PvsAmpl} plots the lightcurve amplitudes versus spin periods
for the 15 KBOs whose lightcurve amplitudes and spin period are known. Open and
filled symbols indicate the KBOs with diameter smaller and larger than
$D=400\,$km, respectively.  Clearly, the smaller and larger KBOs occupy
different regions of the diagram.  For the larger KBOs (black filled circles)
the (small) lightcurve amplitudes are almost independent of the objects' spin
periods.  In contrast, smaller KBOs span a much broader range of lightcurve
amplitudes. Two objects have very low amplitudes: $\SN$ and 1999$\,$KR$_{16}$,
which have diameters $D\sim400\,$km and fall precisely on the boundary of the
two size ranges. The remaining objects hint at a trend of increasing $\Delta m$
with lower spin rates. The one exception is 1999$\,$TD$_{10}$, a Scattered Disk
Object ($e=0.872, a=95.703\,$AU) that spends most of its orbit in rather empty
regions of space and most likely has a different collisional history.

For comparison, Fig.~\ref{Fig.PvsAmpl} also shows results of N-body simulations
of collisions between ``ideal'' rubble piles \citep[gray filled
circles;][]{2000Icar..146..133L}, and the lightcurve amplitude-spin period
relation predicted by ellipsoidal figures of hydrostatic equilibrium
\citep[dashed and dotted lines;][]{1969efe..book.....C,2001Icar..154..432H}.
The latter is calculated from the equilibrium shapes that rotating uniform
fluid bodies assume by balancing gravitational and centrifugal acceleration.
The spin rate-shape relation in the case of uniform fluids depends solely on
the density of the body.  Although fluid bodies behave in many respects
differently from rubble piles, they may, as an extreme case, provide insight on
the equilibrium shapes of gravitationally bound agglomerates.  The lightcurve
amplitudes of both theoretical expectations are plotted assuming an equator-on
observing geometry. They should therefore be taken as upper limits when
compared to the observed KBO amplitudes, the lower limit being zero amplitude.

The simulations of \citeauthor{2000Icar..146..133L}
(\citeyear{2000Icar..146..133L}, hereafter \citetalias{2000Icar..146..133L})
consist of collisions between agglomerates of small spheres meant to simulate
collisions between rubble piles. Each agglomerate consists of $\sim 1000$
spheres, held together by their mutual gravity, and has no initial spin. The
spheres are indestructible, have no sliding friction, and have coefficients of
restitution of $\sim0.8$.  The bulk density of the agglomerates is
2000$\,$kg$\,$m$^{-3}$. The impact velocities range from $\sim$ zero at
infinity to a few times the critical velocity for which the impact energy would
exceed the mutual gravitational binding energy of the two rubble piles. The
impact geometries range from head-on collisions to grazing impacts.  The mass,
final spin period, and shape of the {\em largest remnant} of each collision are
registered (see Table 1 of \citetalias{2000Icar..146..133L}).  From their
results, we selected the outcomes for which the mass of the largest remnant is
equal to or larger than the mass of one of the colliding rubble piles, i.e., we
selected only accreting outcomes.  The spin periods and lightcurve amplitudes
that would be generated by such remnants (assuming they are observed
equator-on) are plotted in Fig.~\ref{Fig.PvsAmpl} as gray circles. Note that,
although the simulated rubble piles have radii of 1$\,$km, since the effects of
the collision scale with the ratio of impact energy to gravitational binding
energy of the colliding bodies \citep{1999Icar..142....5B}, the model results
should apply to other sizes. Clearly, the \citetalias{2000Icar..146..133L}
model makes several specific assumptions, and represents one possible
idealization of what is usually referred to as ``rubble pile''. Nevertheless,
the results are illustrative of how collisions may affect this type of
structure, and are useful for comparison with the KBO data.

The lightcurve amplitudes resulting from the \citetalias{2000Icar..146..133L}
experiment are relatively small ($\Delta m<0.25\,$mag) for spin periods larger
than $P\sim5.5\,$hr (see Fig.~\ref{Fig.PvsAmpl}). Objects spinning faster than
$P=5.5\,$hr have more elongated shapes, resulting in larger lightcurve
amplitudes, up to 0.65 magnitudes. The latter are the result of collisions with
higher angular momentum transfer than the former (see Table 1 of
\citetalias{2000Icar..146..133L}). The maximum spin rate attained by the rubble
piles, as a result of the collision, is $\sim 4.5\,$hr. This is consistent with
the maximum spin expected for bodies in hydrostatic equilibrium with the same
density as the rubble piles ($\rho = 2000\,$kg$\,$m$^{-3}$; see long-dashed
line in Fig.~\ref{Fig.PvsAmpl}). The results of
\citetalias{2000Icar..146..133L} show that collisions between ideal rubble
piles can produce elongated remnants (when the projectile brings significant
angular momentum into the target), and that the spin rates of the collisional
remnants do not extend much beyond the maximum spin permitted to fluid uniform
bodies with the same bulk density.
  
The distribution of KBOs in Fig.~\ref{Fig.PvsAmpl} is less clear.  Indirect
estimates of KBO bulk densities indicate values $\rho\sim1000\,$kg$\,$m$^{-3}$
\citep{2002ARA&A..40...63L}. If KBOs are strengthless rubble piles with such
low densities we would not expect to find objects with spin periods lower than
$P\sim6\,$hr (dashed line in Fig.~\ref{Fig.PvsAmpl}). However, one object
($\CZ$) is found to have a spin period below 5$\,$hr. If this object has a
rubble pile structure, its density must be at least $\sim2000\,$kg$\,$m$^{-3}$.
The remaining 14 objects have spin periods below the expected upper limit,
given their estimated density. Of the 14, 4 objects lie close to the line
corresponding to equilibrium ellipsoids of density $\rho=1000\,$kg$\,$m$^{-3}$.
One of these objects, (20000)$\,$Varuna, has been studied in detail by
\citet{2002AJ....124.1757S}. The authors conclude that (20000)$\,$Varuna is
best interpreted as a rotationally deformed rubble pile with $\rho \le
1000\,$kg$\,$m$^{-3}$.  One object, 2001$\,$QG$_{298}$, has an exceptionally
large lightcurve amplitude ($\Delta m=1.14\,$mag), indicative of a very
elongated shape (axes ratio $a/b>2.85$), but given its modest spin rate
($P=13.8\,$hr) and approximate size ($D\sim 240\,$km) it is unlikely that it
would be able to keep such an elongated shape against the crush of gravity.
Analysis of the lightcurve of this object \citep{2004AJ....127.3023S} suggests
it is a close/contact binary KBO.  The same applies to two other KBOs,
2000$\,$GN$_{171}$ and (33128) 1998$\,$BU$_{48}$, also very likely to be
contact binaries.

To summarize, it is not clear that KBOs have a rubble pile structure, based on
their available rotational properties. A comparison with computer simulations
of rubble pile collisions shows that larger KBOs ($D>400\,$km) occupy the same
region of the period-amplitude diagram as the \citetalias{2000Icar..146..133L}
results.  This is not the case for most of the smaller KBOs ($D\le 400\,$km),
which tend to have larger lightcurve amplitudes for similar spin periods. If
most KBOs are rubble piles then their spin rates set a lower limit to their
bulk density: one object ($\CZ$) spins fast enough that its density must be at
least $\rho\sim2000\,$kg$\,$m$^{-3}$, while 4 other KBOs (including
(20000)$\,$Varuna) must have densities larger than
$\rho\sim1000\,$kg$\,$m$^{-3}$.  A better assessment of the inner structure of
KBOs will require more observations, and detailed modelling of the collisional
evolution of rubble-piles.

\section{Conclusions}

We have collected and analyzed R-band photometric data for 10 Kuiper Belt
objects, 5 of which have not been studied before. No significant brightness
variations were detected from KBOs $\CM$, $\RZ$, $\TS$.  Previously observed
KBOs $\WH$, $\TC$, and $\EB$ were confirmed to have very low amplitude
lightcurves ($\Delta m \le 0.1\,$mag).  $\SN$, $\DF$, and $\CZ$ were shown to
have periodic brightness variations. Our lightcurve amplitude statistics are
thus: 3 out of 10 (30\%) observed KBOs have $\Delta m \ge 0.15\,$mag, and 1 out
of 10 (10\%) has $\Delta m \ge 0.40\,$mag. This is consistent with previously
published results.

The rotational properties that we obtained were combined with existing data in
the literature and the total data set was used to investigate the distribution
of spin period and shapes of KBOs. Our conclusions can be summarized as
follows:
\begin{enumerate}
\item KBOs with diameters $D>200\,$km have a mean spin period of $9.23\,$hr,
and thus rotate slower on average than main belt asteroids of similar size
($\langle P \rangle_{\rm MBAs}=6.48\,$hr). The probability that the two
distributions are drawn from the same parent distribution is 0.7\%, as judged
by the KS test.
\item 26 of 37 KBOs (70\%, $D>200\,$km) have lightcurve amplitudes below
$0.15\,$mag. In the asteroid belt only 10 of the 27 (37\%) asteroids in the
same size range have such low amplitude lightcurves. This difference is
significant at the 99.2\% level according to the Fisher exact test.
\item KBOs with diameters $D>400\,$km have lightcurves with significantly
(98.5\% confidence) smaller amplitudes ($\langle \Delta m \rangle=0.13\,$mag,
$D>400\,$km) than KBOs with diameters $D\le 400\,$km ($\langle \Delta m
\rangle=0.25\,$mag, $D\le 400\,$km).
\item These two size ranges seem to have different shape distributions, but the
few existing data do not render the difference statistically significant. Even
though the shape distributions in the two size ranges are not inconsistent, the
best-fit power-law solutions predict a larger fraction of round objects in the
$D>400\,$km size range ($f(a/b<1.2)\sim70^{+12}_{-19}\%$) than in the group of
smaller objects ($f(a/b<1.2)\sim42^{+20}_{-15}\%$).
\item The current KBO lightcurve data are too sparse to allow a conclusive
assessment of the inner structure of KBOs.
\item KBO $\CZ$ has a spin period of $P=4.71\,$hr. If this object has a rubble
pile structure then its density must be $\rho \gtrsim 2000\,$kg$\,$m$^{-3}$. If
the object has a lower density then it must have internal strength.
\end{enumerate}

The analysis presented in this paper rests on the assumption that the available
sample of KBO rotational properties is homogeneous.  However, in all likelihood
the database is biased. The most likely bias in the sample comes from
unpublished flat lightcurves. If a significant fraction of flat lightcurves
remains unreported then points 1 and 2 above could be strengthened, depending
on the cause of the lack of brightness variation (slow spin or round shape).
On the other hand, points 3 and 4 could be weakened if most unreported cases
correspond to smaller KBOs. Better interpretation of the rotational properties
of KBOs will greatly benefit from a larger and more homogeneous dataset.

\acknowledgments

This work was supported by grants from the Netherlands Foundation for Research
(NWO), the Leids Kerkhoven-Bosscha Fonds (LKBF), and a NASA grant to D. Jewitt.
We are grateful to Scott Kenyon, Ivo Labb\'e, and D. J. for helpful discussion
and comments.


\end{document}